\documentclass[APA,STIX1COL]{WileyNJD-v2}
\usepackage{siunitx}
\usepackage{graphicx}
\usepackage{subcaption}
\usepackage{float}        
\usepackage{threeparttable}

\usepackage{soul}

\graphicspath{
  {figures/manuscript/}
  {Figures/manuscript/}
  {figures/appendix_cutoff_panels/}
  {Figures/appendix_cutoff_panels/}
  {figures/multivariate_synthesis_by_cutoff/}
  {Figures/multivariate_synthesis_by_cutoff/}
}

\articletype{Research Article} 

\received{Date Month Year}
\revised{Date Month Year}
\accepted{Date Month Year}
\copyyear{2025}
\startpage{1}

\begin{document}

\title{Bayesian Quantile-Based Correction and Synthesis of Hydrologic Products}

\author[1]{Antonio De Leon*}
\author[1]{Raquel Prado}
\author[1]{Bruno Sans\'o}

\authormark{De Leon \textsc{et al.}} 

\address[1]{\orgdiv{Department of Statistics}, \orgname{University of California, Santa Cruz}, \orgaddress{\state{California}, \country{USA}}}

\corres{*Corresponding author: Antonio De Leon. \email{jaguir26@ucsc.edu}}

\presentaddress{University of California, Santa Cruz, Santa Cruz, CA, USA.}

\abstract[Summary]{River-flow forecasting requires predictive distributions that remain informative in both routine and extreme conditions. We develop a Bayesian quantile-based correction-and-synthesis framework built on Dynamic Quantile Linear Models (DQLMs). The framework links U.S. Geological Survey (USGS) observations, retrospective products, and ensemble forecast products through a shared latent quantile process, learns dynamic discrepancies for each external source, and combines quantile-specific posterior predictions into a single predictive distribution. We also adapt variational Bayes inference to the extended dynamic quantile linear model using Laplace--Delta approximations for non-conjugate parameters. The methodology is illustrated using daily flow for the San Lorenzo River together with products from the European Centre for Medium-Range Weather Forecasts (ECMWF) Global Flood Awareness System (GloFAS) and the National Oceanic and Atmospheric Administration (NOAA) National Weather Service (NWS), with emphasis on medium-range forecasting and uncertainty quantification across multiple quantile levels.}


\jnlcitation{\cname{%
\author{De Leon A.},
\author{Prado R.}, and
\author{Sans\'o B.}} (\cyear{2025}),
\ctitle{Bayesian quantile-based correction and synthesis of hydrologic products}, \cjournal{Environmetrics}, \cvol{XX:1--20}.}

\maketitle

\noindent\textbf{Abbreviations:} CRPS, Continuous Ranked Probability Score; DQLM, Dynamic Quantile Linear Model.

\section{INTRODUCTION}

River flow forecasts are essential for flood preparedness, drought monitoring, reservoir operations, and water-supply planning. In regions such as Santa Cruz, California, forecast accuracy depends on both hydrological and meteorological information. These two uncertainty sources are related but distinct. Hydrological uncertainty arises from model structure, parameters, states, and observations, whereas meteorological uncertainty enters through imperfect precipitation and related atmospheric forcing fields \citep{tarek2021daily, thiboult2016accounting, kim2019hybrid, haiden2021evaluation}. A useful statistical forecasting framework should keep these roles clear while accounting for their interaction in the final predictive distribution.

Operational hydrological forecasts are commonly produced with conceptual or physically based models. Conceptual formulations remain especially practical for prediction because their lower-dimensional structure is often easier to specify, calibrate, and deploy operationally. Forecasts from these systems are increasingly distributed as ensemble products, which are useful but often differ in horizon, spatial resolution, assimilation strategy, and uncertainty representation \citep{molteni1996ecmwf, isaksen2010ensemble, demargne2013science, barlage2021importance, johnson2023comprehensive}. Retrospective products introduce a different source of information: they provide historically consistent reconstructions that can be compared against observations to learn systematic discrepancies. Operational hydrologic centers therefore provide both retrospective analysis products, which support hindcast-style evaluation and discrepancy learning, and forecast ensembles, which provide origin-time information about future flow conditions. The statistical problem addressed here is how to combine observations, retrospective products, and forecast products in a way that improves predictive distributions for river flow.

This setting is closely related to forecast combination and statistical post-processing, where multiple predictive sources are combined to improve calibration and sharpness \citep{siegert2019forecast, vannitsem, wang2022forecastcombinations50yearreview}. Bayesian formulations provide a coherent way to propagate uncertainty and update predictions as new information becomes available \citep{duan2006bma, yumnam2022quantile, aastveit2022quantifying, du2018ensemble_forecasting, doubleday, kleiber2011geostatistical, mcalinn2019dynamic, tallman2023bayesianpredictivedecisionsynthesis, johnson2023bayesian}. Our focus is on a quantile-based version of this problem because river-flow forecasts must remain informative in both routine and extreme conditions.

We develop a Bayesian quantile-based correction-and-synthesis framework for river flow forecasts. The framework builds on Dynamic Quantile Linear Models (DQLMs) \citep{goncalves, gourieroux2008dqlm, barata2021_flex_quantile} and has three main components. First, we extend posterior inference for the model using scalable variational Bayes methods for the extended dynamic quantile linear model, following the Laplace--Delta strategy of \citet{wang2013nonconjugatevb}. Second, we use retrospective products to learn dynamic discrepancies relative to U.S. Geological Survey (USGS) observations and propagate those discrepancies into the forecast period. Third, we synthesize the resulting quantile-specific predictive distributions into one predictive distribution for the river flow process.

\begin{figure*}[hbt!]
\centering
\centerline{\includegraphics[width=450pt]{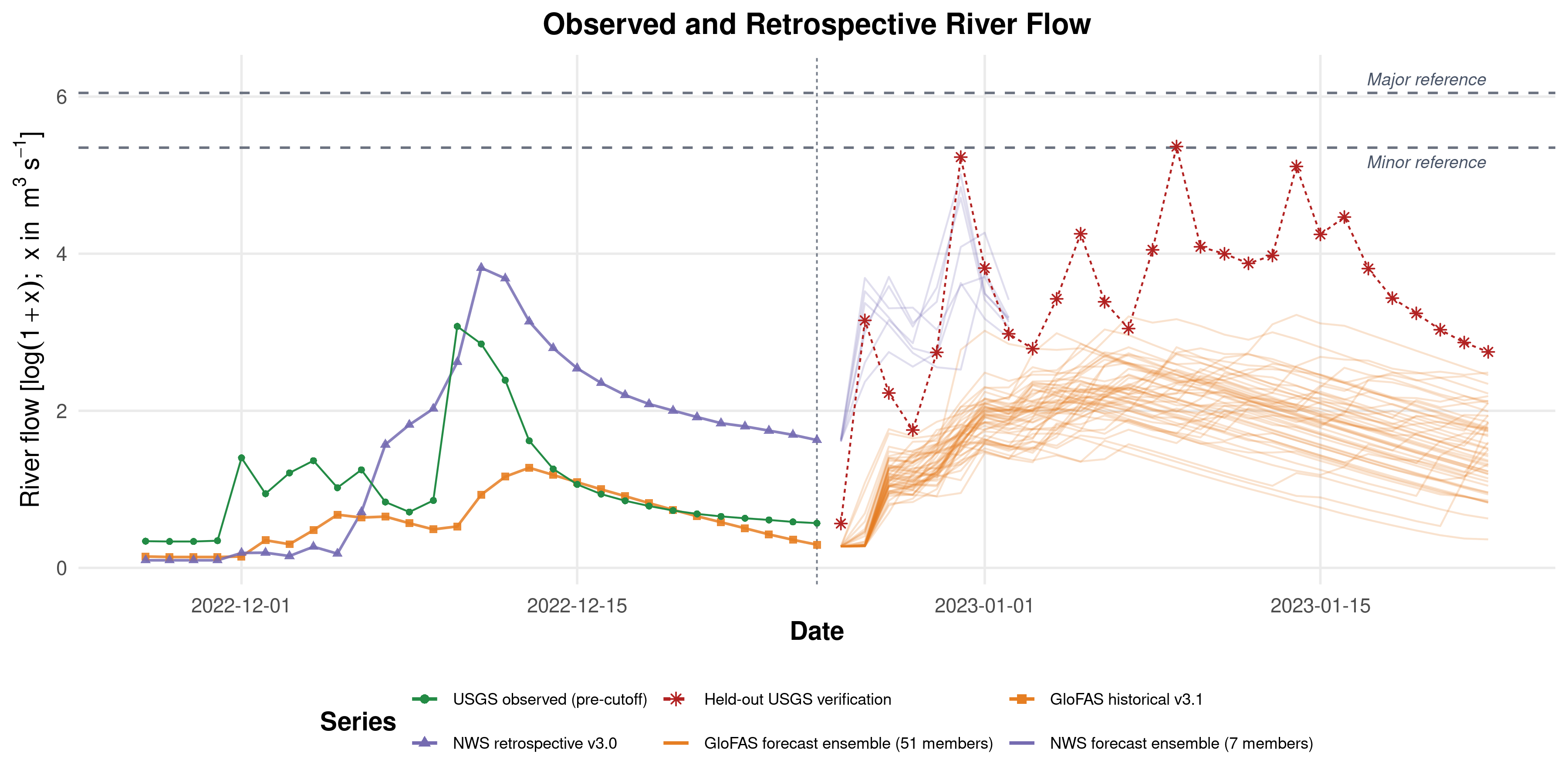}}
\caption{
Representative forecast-origin display for the December 25, 2022 cutoff, spanning November 27, 2022 through January 22, 2023 on the $\log(1+x)$ scale with $x$ in m$^3$/s. Before the cutoff, the panel shows USGS observations and the NWS and GloFAS retrospective products used for source-discrepancy learning; after the cutoff, it shows held-out USGS verification together with the operational NWS and GloFAS forecast ensembles. Horizontal dashed lines show approximate current-rating discharge equivalents of the current NWS minor and major stage categories at BTEC1/USGS 11160500; they are included as operational magnitude references rather than as historical flood-stage classifications.
}
\label{fig:ensembles}
\end{figure*}

We study this framework using daily river flow for the San Lorenzo River in California, together with retrospective and forecast products from the European Centre for Medium-Range Weather Forecasts (ECMWF) Global Flood Awareness System (GloFAS) and the National Oceanic and Atmospheric Administration (NOAA) National Weather Service (NWS). These products are operationally useful but not directly exchangeable: they differ in horizon, update frequency, spatial support, and uncertainty representation, so a statistical synthesis must account for those differences rather than treat the inputs as homogeneous. Figure~\ref{fig:ensembles} illustrates the information structure at one representative rolling-origin cutoff, December 25, 2022: before the cutoff, observations and retrospective products are available for discrepancy learning; after the cutoff, issued forecast ensembles provide source-specific information about future flow. The empirical focus is forecasting performance and uncertainty quantification across multiple quantile levels, rather than only historical fit or methodological development in isolation. Although the application is hydrological, the same structure is relevant in other settings where observations, retrospective products, and forecast products all play distinct roles.

The remainder of the paper is organized as follows. Section~\ref{sec:methodology} presents the modeling framework and inference strategy. Section~\ref{sec:data} describes the San Lorenzo River application, the data sources, and the rolling-origin forecasting design. Section~\ref{sec:forecastvalidation} reports the out-of-sample forecast validation results, and Section~\ref{sec:interpretation} presents supporting interpretation for the selected specification. Section~\ref{sec5} discusses conclusions and further work.

\section{METHODOLOGY}
\label{sec:methodology}

Throughout the article, we use bold uppercase letters (e.g., $\mathbf{W}, \mathbf{F}, \mathbf{G}$) to denote matrices, and bold lowercase letters (e.g., $\mathbf{x}, \boldsymbol{\psi}, \boldsymbol{\theta}$) to denote vectors. We denote the identity matrix of dimension $r \times r$ as $\mathbf{I}_r$, a vector of zeros of dimension $r \times 1$ as $\mathbf{0}_r$, and a vector of ones of dimension $r \times 1$ as $\mathbf{1}_r$.

\subsection{Extended Asymmetric Laplace Likelihood}
\label{subsec:exal}

The use of the Asymmetric Laplace (AL) distribution for quantile inference within a Bayesian framework is well-established with both empirical and theoretical justifications \citep{yu2001bayesian, sriram2013theoretical}. We use the extended asymmetric Laplace (exAL) distribution, which addresses limitations of the AL working likelihood, particularly its restricted skewness. \citet{yan2025new} proposed the exAL distribution, $\text{exAL}_{p_0}(\mu, \sigma, \gamma)$, where $\mu$ represents the $p_0$th quantile of interest, $\sigma$ is a positive scale parameter, and $\gamma$ is the skewness parameter. The parameter $\gamma$ has bounded support over the interval $(L, U)$, where $L$ is the negative root of $g(\gamma) = 1 - p_0$ and $U$ is the positive root of $g(\gamma) = p_0$, with $g(\gamma) = 2 \Phi(-|\gamma|) \exp(\gamma^2 / 2)$ and $\Phi$ denoting the standard normal cumulative distribution function. For computational implementation, we use the extended stochastic representation of an exAL random variable described in \citet{yan2025new}. That is, if $y \sim \text{exAL}(\mu, \sigma, \gamma)$, then there exist independent random variables $\epsilon \sim \mathcal{N}(0, 1)$, $z \sim \text{Exp}(1)$, and $s \sim \mathcal{N}^+(0, 1)$, (where $\mathcal{N}^+(0, 1)$ denotes the standard normal distribution truncated to the positive real line), such that  \( y = \mu + C(\gamma; p_0)\sigma |\gamma| s + A(\gamma; p_0) \sigma z + \sqrt{B(\gamma; p_0) \sigma^2 z} \cdot \epsilon, \) where $A(\gamma; p_0) = \frac{1 - 2 p(\gamma; p_0)}{p(\gamma; p_0) \, [1 - p(\gamma; p_0)]}$, $B(\gamma; p_0) = \frac{2}{p(\gamma; p_0) \, [1 - p(\gamma; p_0)]}$, $C(\gamma; p_0) = [I(\gamma > 0) - p(\gamma; p_0)]^{-1}$, and $p(\gamma; p_0) = I(\gamma < 0) + \{p_0 - I(\gamma < 0)\} / g(\gamma)$.

\subsection{State-Space Model Specification}
\label{subsec:fullmodel}

For a fixed target quantile $p_0$, we use a single state-space model whose latent process evolves across both the pre-cutoff observation period and the forecast horizon. The cutoff, denoted by \(T\), separates the data used for fitting from the post-cutoff observations reserved for verification. The latent river-flow quantile is represented by a trend-and-seasonal state \(\boldsymbol{\theta}_t \in \mathbb{R}^p\), observed through the loading vector \(\mathbf{F}_t\). Each external product family \(j\) has its own discrepancy state \(\boldsymbol{\delta}_t^j \in \mathbb{R}^p\), allowing that product to depart systematically from the observed river-flow series. A scalar transfer adjustment \(\zeta_t\) incorporates exogenous covariates \(\mathbf{x}_t \in \mathbb{R}^m\), whose time-varying effects are collected in \(\boldsymbol{\psi}_t \in \mathbb{R}^m\). Let \(\mathcal{J}=\{1,\ldots,J\}\) denote the set of external product families. For each \(j\in\mathcal{J}\), let \(\mathcal{I}_j=\{1,\ldots,I_j\}\) index forecast members and \(\mathcal{K}_j=\{1,\ldots,K_{(j)}\}\) index forecast lead times. Finally, define \(K_{\max}=\max_{j\in\mathcal{J}}K_{(j)}\), the maximum forecast horizon across all sources.

The expanded specification is:
\begin{alignat}{3}
\text{Observed measurements:}\quad&
y_t^o
&\sim{}& \text{exAL}_{p_0}\!\left(\mathbf{F}_t' \boldsymbol{\theta}_t + \zeta_t, \sigma^0, \gamma^0\right),
&\qquad& 1 \le t \le T, \label{eq:unified_usgs}
\\
\text{Retrospective products:}\quad&
z_t^j
&\sim{}& \text{exAL}_{p_0}\!\left(\mathbf{F}_t'(\boldsymbol{\theta}_t + \boldsymbol{\delta}_t^j) + \zeta_t, \sigma^j, \gamma^j\right),
&\qquad& j \in \mathcal{J},\ 1 \le t \le T, \label{eq:unified_retro}
\\
\text{Forecast products:}\quad&
y_T^{j,i}(k)
&\sim{}& \text{exAL}_{p_0}\!\left(\mathbf{F}_{T+k}'(\boldsymbol{\theta}_{T+k} + \boldsymbol{\delta}_{T+k}^j) + \zeta_{T+k}, \sigma^j, \gamma^j\right),
&\qquad& j \in \mathcal{J},\ i \in \mathcal{I}_j,\ k \in \mathcal{K}_j, \notag
\\
\text{Trend and seasonal states:}\quad&
\boldsymbol{\theta}_t \mid \boldsymbol{\theta}_{t-1}
&\sim{}& \mathcal{N}\!\left(\mathbf{G}_t \boldsymbol{\theta}_{t-1}, \mathbf{W}_t^\theta\right),
&\qquad& 1 \le t \le T+K_{\max}, \label{eq:unified_theta}
\\
\text{Discrepancy states:}\quad&
\boldsymbol{\delta}_t^j \mid \boldsymbol{\delta}_{t-1}^j
&\sim{}& \mathcal{N}\!\left(\mathbf{G}_t \boldsymbol{\delta}_{t-1}^j, \mathbf{W}_t^{\delta^j}\right),
&\qquad& j \in \mathcal{J},\ 1 \le t \le T+K_{\max}, \label{eq:unified_delta}
\\
\text{Transfer adjustment:}\quad&
\zeta_t \mid \zeta_{t-1}, \boldsymbol{\psi}_{t-1}
&\sim{}& \mathcal{N}\!\left(\lambda \zeta_{t-1} + \mathbf{x}_t'\boldsymbol{\psi}_{t-1}, w_t^\zeta\right),
&\qquad& 1 \le t \le T+K_{\max}, \label{eq:unified_transfer}
\\
\text{Covariate-effect states:}\quad&
\boldsymbol{\psi}_t \mid \boldsymbol{\psi}_{t-1}
&\sim{}& \mathcal{N}\!\left(\boldsymbol{\psi}_{t-1}, \mathbf{W}_t^\psi\right),
&\qquad& 1 \le t \le T+K_{\max}. \notag
\end{alignat}

Here \(\mathbf{F}_t \in \mathbb{R}^{p}\), \(\mathbf{G}_t \in \mathbb{R}^{p\times p}\), \(\mathbf{W}_t^\theta \in \mathbb{R}^{p\times p}\), \(\mathbf{W}_t^{\delta^j} \in \mathbb{R}^{p\times p}\), \(\mathbf{W}_t^\psi \in \mathbb{R}^{m\times m}\), and \(w_t^\zeta \in \mathbb{R}_+\) is the scalar evolution variance for the transfer state \(\zeta_t\). The parameter \(\lambda \in \mathbb{R}\) governs persistence in the transfer adjustment, while \(\mathbf{x}_t'\boldsymbol{\psi}_{t-1}\) is the covariate contribution at time \(t\). During the forecast window, \(\mathbf{x}_{T+k}\) denotes the forecast-window covariate vector available from the staged cutoff-specific input bundle. The notation \(y_T^{j,i}(k)\) denotes the \(k\)-step-ahead forecast issued at time \(T\) by member \(i\) of source \(j\).

The first two observation equations correspond to the pre-cutoff observation window, where the available channels are the observed USGS flow and the retrospective products. The third observation equation corresponds to the forecast window, where issued forecast products enter through the same source-specific discrepancy structure. The state-evolution equations define the roles of the dynamic components: \(\boldsymbol{\theta}_t\) carries the river-flow level and seasonal structure, \(\boldsymbol{\delta}_t^j\) captures systematic source-specific departures from the observations, and the transfer component maps exogenous covariates into the quantile process through \(\zeta_t\) and \(\boldsymbol{\psi}_t\).

Define
\[
\boldsymbol{\omega}_t' = [\zeta_t,\ \boldsymbol{\psi}_t'], \qquad
\boldsymbol{\eta}_t' = [\boldsymbol{\theta}_t',\ (\boldsymbol{\delta}_t^1)',\ \ldots,\ (\boldsymbol{\delta}_t^J)',\ \boldsymbol{\omega}_t'],
\]
where \(\boldsymbol{\omega}_t \in \mathbb{R}^{m+1}\) and \(\boldsymbol{\eta}_t \in \mathbb{R}^{q}\), with \(q = (J+1)p + m + 1\). The transfer update can be written as
\[
\mathbf{G}^{\text{trans}}_t =
\begin{bmatrix}
\lambda & \mathbf{x}_t' \\
\mathbf{0}_m & \mathbf{I}_m
\end{bmatrix}.
\]
Let \(\mathcal{J}_0=\{0\}\cup\mathcal{J}\), where \(s=0\) denotes the USGS observation channel and \(s\in\mathcal{J}\) denotes an external product family. Let \(y_t^0 \equiv y_t^o\) denote the observation channel and \(y_t^s \equiv z_t^s\), \(s\in\mathcal{J}\), denote the retrospective-product channels. The corresponding compact form is:
\begin{alignat}{3}
\text{Pre-cutoff observations:}\quad&
y_t^s \mid \boldsymbol{\eta}_t, \sigma^s, \gamma^s
&\sim{}& \text{exAL}_{p_0}\!\left(\mathbf{h}_{t,s}' \boldsymbol{\eta}_t, \sigma^s, \gamma^s\right),
&\qquad& 1 \le t \le T,\ s \in \mathcal{J}_0, \label{eq:compact_est_obs}
\\
\text{Pre-cutoff state evolution:}\quad&
\boldsymbol{\eta}_t \mid \boldsymbol{\eta}_{t-1}
&\sim{}& \mathcal{N}\!\left(\boldsymbol{D}_t \boldsymbol{\eta}_{t-1}, \boldsymbol{\Lambda}_t\right),
&\qquad& 1 \le t \le T, \label{eq:compact_est_state}
\\
\text{Forecast observations:}\quad&
y_T^{j,i}(k) \mid \boldsymbol{\eta}_{T+k}, \sigma^j, \gamma^j
&\sim{}& \text{exAL}_{p_0}\!\left(\mathbf{e}_{T+k,j}' \boldsymbol{\eta}_{T+k}, \sigma^j, \gamma^j\right),
&\qquad& j \in \mathcal{J},\ i \in \mathcal{I}_j,\ k \in \mathcal{K}_j, \label{eq:compact_fc_obs}
\\
\text{Forecast state evolution:}\quad&
\boldsymbol{\eta}_{T+k} \mid \boldsymbol{\eta}_{T+k-1}
&\sim{}& \mathcal{N}\!\left(\boldsymbol{M}_{T+k} \boldsymbol{\eta}_{T+k-1}, \mathbf{W}_{T+k}\right),
&\qquad& 1 \le k \le K_{\max}. \label{eq:compact_fc_state}
\end{alignat}

Here
\[
\big(\mathbf{F}_t^{\text{trans}}\big)' = [1,\ \mathbf{0}_m'], \qquad
\mathbf{h}_{t,0}' =
\big[\mathbf{F}_t',\ \mathbf{0}_{pJ}',\ \big(\mathbf{F}_t^{\text{trans}}\big)'\big],
\qquad
\mathbf{h}_{t,j}' =
\big[\mathbf{F}_t',\ \mathbf{0}_{p(j-1)}',\ \mathbf{F}_t',\ \mathbf{0}_{p(J-j)}',\ \big(\mathbf{F}_t^{\text{trans}}\big)'\big],
\quad j \in \mathcal{J},
\]
with \(\mathbf{F}_t^{\text{trans}} \in \mathbb{R}^{m+1}\), \(\mathbf{h}_{t,0}, \mathbf{h}_{t,j} \in \mathbb{R}^{q}\), and
\[
\boldsymbol{D}_t = \text{BlockDiag}\!\left(\mathbf{G}_t,\ \underbrace{\mathbf{G}_t,\ldots,\mathbf{G}_t}_{J\ \text{discrepancy blocks}},\ \mathbf{G}^{\text{trans}}_t\right), \qquad
\boldsymbol{\Lambda}_t = \text{BlockDiag}\!\left(\mathbf{W}^{\theta}_t,\ \underbrace{\mathbf{W}^{\delta^1}_t,\ldots,\mathbf{W}^{\delta^J}_t}_{J\ \text{source-specific discrepancy blocks}},\ \text{BlockDiag}\!\left(w_t^{\zeta}, \mathbf{W}^{\psi}_t\right)\right),
\]
where \(\boldsymbol{D}_t, \boldsymbol{\Lambda}_t \in \mathbb{R}^{q\times q}\) and \(\sigma^s \in \mathbb{R}_+\), \(\gamma^s \in \mathbb{R}\) denote the channel-specific scale and skewness parameters for \(s \in \mathcal{J}_0\). Thus, \(\mathbf{h}_{t,0}' \boldsymbol{\eta}_t = \mathbf{F}_t' \boldsymbol{\theta}_t + \zeta_t\) and \(\mathbf{h}_{t,j}' \boldsymbol{\eta}_t = \mathbf{F}_t'(\boldsymbol{\theta}_t + \boldsymbol{\delta}_t^j) + \zeta_t\). For the forecast window, \(\mathbf{e}_{T+k,j}\) is the analogue of \(\mathbf{h}_{t,j}\): it extracts the latent trend-and-seasonal block, the discrepancy block for source \(j\), and the transfer block. The matrices \(\boldsymbol{M}_{T+k}\) and \(\mathbf{W}_{T+k}\) preserve the same block ordering while propagating the active forecast-window state.

This compact representation highlights the conditionally Gaussian state-space structure induced by the stochastic representation in Subsection~\ref{subsec:exal}. The observation equations are conditionally Gaussian given the latent augmentation variables, and the state evolution remains Gaussian. The model can therefore be handled with standard state-space methods, including filtering, smoothing, and sampling recursions.

\subsection{Prior Specification and Discounting}
\label{subsec:prioranddisc}
To complete the model we specify prior distributions for the initial states, the scale, and skewness parameters.
Following \citet{barata2021_flex_quantile}, for each channel-specific skewness parameter, we assign a truncated Student-\(t\) prior:
\[
\gamma^s \sim t_{(L,U)}(0,\phi^s,\nu^s), \qquad s\in\mathcal{J}_0,
\]
where \(\phi^s\) and \(\nu^s\) denote the scale and degrees of freedom hyperparameters, respectively. For the channel-specific scale parameters, we use inverse-gamma priors,
\[
\sigma^s \sim IG(a_{\sigma^s}, b_{\sigma^s}), \qquad s\in\mathcal{J}_0,
\]
with \(a_{\sigma^s}\) and \(b_{\sigma^s}\) as shape and scale hyperparameters. The full latent state is initialized as \(\boldsymbol{\eta}_0 \sim \mathcal{N}(\mathbf{m}_0, \mathbf{C}_0)\), with \(\mathbf{m}_0 \in \mathbb{R}^{q}\) and \(\mathbf{C}_0 \in \mathbb{R}^{q\times q}\). This initial prior covers the shared trend-and-seasonal quantile block, the source-specific discrepancy blocks, and the transfer block.

Discount factors \citep{prado2021time} are used for adaptive evolution of the covariance matrices at the state level. Discount factors lie in the $(0,1]$ range, with 1 indicating no dynamic evolution and smaller values implying greater temporal adaptability. We use component discounting. The application-specific choices are given in Subsection~\ref{subsec:specification}. During the forecast window ($t \in \{T+1, \ldots, T+K_{\max}\}$), we set the Inverse Wishart prior parameters as $\mathbf{S}_t = (\nu_t - p_t - 1)\, c\, \mathbf{W}_T$ and $\nu_t = (p_t + 1 + \epsilon)$, with $\epsilon, c > 0$, where \(p_t\) denotes the dimension of the active forecast-window state at time \(t\), \(\mathbf{S}_t \in \mathbb{R}^{p_t\times p_t}\), and \(\mathbf{W}_T \in \mathbb{R}^{p_t\times p_t}\) after restricting the full pre-cutoff covariance to the active forecast-window block. Under this specification, the posterior mean is
\(
\mathbb{E}[\mathbf{W}_t \mid \mathcal{D}_t] = \frac{\epsilon}{1 + \epsilon}(c\, \mathbf{W}_T) + \frac{1}{1 + \epsilon} \mathbb{E}\left[ (\boldsymbol{\eta}_t - \mathbf{M}_t \boldsymbol{\eta}_{t-1}) (\boldsymbol{\eta}_t - \mathbf{M}_t \boldsymbol{\eta}_{t-1})' \mid \mathcal{D}_t \right],
\)
where $\mathcal{D}_t$ denotes all data available up to and including time $t$.

This posterior mean is a convex combination of two terms: the covariance learned before the cutoff, carried forward through \(c\mathbf{W}_T\), and the forecast-window residual covariance implied by the propagated state. The scalar \(c\) controls the scale of the carried-forward covariance anchor, while \(\epsilon\) controls how strongly that anchor is retained. In the reported analysis we use this prior as a short-horizon stabilization device: it prevents the forecast-window covariance from being driven entirely by a small number of post-cutoff forecast-product updates, while still allowing the forecast-window state evolution to adapt through the second term in the convex combination.
Further details regarding prior specification and discount factors are provided in Subsection~\ref{subsec:specification}.

\subsection{Posterior Computation}
\label{subsec:postinf}
Initially, we explored the posterior distribution of all model parameters using a Markov chain Monte Carlo (MCMC) algorithm.
Algorithm~\ref{alg1} gives the core MCMC updates in a compact source-indexed notation. As indicated by Equation~\ref{eq:compact_est_obs}, during the pre-cutoff observation window the response variables are the USGS observations and retrospective products. After the cutoff, Equation~\ref{eq:compact_fc_obs} applies and the response variables correspond to the forecast products. Careful bookkeeping is needed because forecast horizons differ across agencies. Gibbs sampling is facilitated by the fact that all full posterior conditionals are available in closed form except for the channel-specific skewness parameters \(\gamma^s\), \(s\in\mathcal{J}_0\). For these, we use a Metropolis--Hastings random walk on an unconstrained reparameterization. Posterior sampling of the state parameters relies on a modified Forward Filtering Backward Sampling (FFBS) algorithm that leverages their conditional normality (see Algorithm~\ref{alg:Alg2}).


Given the computational demands of traditional Markov Chain Monte Carlo (MCMC) methods on large datasets, we also implement a Mean Field Variational Bayes (MFVB) algorithm to approximate posterior inference.
\citet{barata2021_flex_quantile} proposed a Variational Bayes algorithm that embeds an importance sampling scheme for non-conjugate parameters; however, this scheme risks particle collapse in high dimensions and requires further constraints to avoid degeneracy. In particular, it fixes the scale parameter of the exAL using pre-estimated values from an AL model.
Building on \citet{barata2021_flex_quantile}, we address stability issues in their importance sampling approach by adopting the Variational Bayes Laplace-Delta (VB-LD) methodology from \citet{wang2013nonconjugatevb}. This framework replaces the importance sampling strategy to handle non-conjugate parameters with a dual strategy: Laplace approximations for non-conjugate parameters and the use of the Delta method for computing expectations of functionals of the non-conjugate parameters.

To illustrate our VB-LD strategy, consider a generic joint distribution \( p(D,\boldsymbol{\alpha}, \boldsymbol{\beta}) \), where \(D\) denotes the data, \(\boldsymbol{\beta}\) denotes conjugate quantities, and \(\boldsymbol{\alpha}\) represents non-conjugate parameters. We employ a Mean Field Variational Bayes (MFVB) approximation, assuming a factorized posterior \(q(\boldsymbol{\alpha})q(\boldsymbol{\beta})\), and optimize these approximations via Coordinate Ascent Variational Inference (CAVI). The CAVI framework in this setting iteratively updates two components: the conjugate quantities \(\boldsymbol{\beta}\) through \(q(\boldsymbol{\beta}) \propto \exp\{\mathbb{E}_{q(\boldsymbol{\alpha})}[\log p(D,\boldsymbol{\alpha},\boldsymbol{\beta})]\}\), which admits closed-form solutions, and the non-conjugate parameters \(\boldsymbol{\alpha}\) via \(q(\boldsymbol{\alpha}) \propto \exp\{\ell(\boldsymbol{\alpha})\}\), where \(\ell(\boldsymbol{\alpha})=\mathbb{E}_{q(\boldsymbol{\beta})}[\log p(D,\boldsymbol{\alpha},\boldsymbol{\beta})]\). The latter update is intractable due to its non-analytic normalizing constant, which we address through two steps. First, a Laplace approximation constructs a Gaussian surrogate \(q(\boldsymbol{\alpha}) \approx \mathcal{N}(\hat{\boldsymbol{\alpha}},\Sigma)\), where \(\hat{\boldsymbol{\alpha}}=\arg\max_{\boldsymbol{\alpha}}\ell(\boldsymbol{\alpha})\) and \(\Sigma=-[\nabla^2\ell(\hat{\boldsymbol{\alpha}})]^{-1}\), matching the mode and curvature of \(\ell(\boldsymbol{\alpha})\). Second, expectations \(\mathbb{E}_{q(\boldsymbol{\alpha})}[g(\boldsymbol{\alpha})]\) required for updating \(\boldsymbol{\beta}\) are approximated via a second-order Taylor expansion (Delta Method): \(\mathbb{E}[g(\boldsymbol{\alpha})]\approx g(\hat{\boldsymbol{\alpha}})+\frac{1}{2}\text{Tr}\{\nabla^2 g(\hat{\boldsymbol{\alpha}})\Sigma\}\), propagating uncertainty through \(\Sigma\) without stochastic sampling. The VB-LD algorithm iterates between Laplace mode-finding (e.g., via Newton-Raphson) and conjugate updates using these Delta-approximated moments until the measured Evidence Lower Bound (ELBO) converges.
For reproducibility, implementation pseudocode for the VB algorithm is provided in Algorithm~\ref{alg:Alg3} and Algorithm~\ref{alg:Alg4}.

Our approach fits a separate state-space model for each target quantile level, which allows the seven quantile-specific models used in the application to be fitted in parallel. In the San Lorenzo application, the full quantile set can be refit on operational time scales once observations, retrospective products, forecast products, and forecast covariates have been staged; representative publication runs take about two hours end-to-end on a Linux server with 64 cores and 503 GiB of RAM. These timings use the practical runtime fields \texttt{runtime\_sec\_total} and \texttt{runtime\_sec}; they are combined wall-clock measurements because fitting and forecasting were not timed separately, and they are hardware- and implementation-dependent. The main computational cost arises from repeated filtering and smoothing operations and increases with the length of the time series and the dimension of the latent state.

\subsection{Posterior Predictive Synthesis Across Quantiles}
\label{subsec:quantilesynthesis}

The model is fitted separately at a finite set of target quantile levels, but forecast validation and visualization require a single predictive distribution at each time point. Because independently fitted quantile models do not automatically define one monotone predictive quantile function, we apply a post-fit synthesis step to the posterior predictive output from the fitted quantile-specific models. This step uses only posterior predictive output from the fitted models: it does not refit the state-space models, alter the VB updates, or use held-out post-cutoff USGS observations.

Let \(0<\tau_1<\cdots<\tau_L<1\) denote the fitted quantile levels and let \(Y_{\ell,t}^{(r)}\) be posterior predictive draws from the model fitted at level \(\tau_\ell\), for \(r=1,\ldots,R\). Let \(q_{\ell,t}\) be the empirical \(\tau_\ell\)-quantile of those draws. Because the \(L\) quantile models are fitted independently, the anchor values \(q_{\ell,t}\) are first adjusted with isotonic regression \citep{barlow1972},
\[
\mathbf{m}_t^*
= \underset{m_1 \leq \cdots \leq m_L}{\arg\min}
\sum_{\ell=1}^L (q_{\ell,t}-m_\ell)^2 .
\]
The posterior draws from each quantile lane are then shifted to match the adjusted anchor,
\[
Y_{\ell,t}^{(r),\mathrm{adj}}
= Y_{\ell,t}^{(r)} + (m_{\ell,t}^* - q_{\ell,t}).
\]
The adjusted quantile anchors define an initial synthesized quantile curve by interpolation across adjacent fitted levels. We evaluate this curve on a dense probability grid and apply monotone rearrangement \citep{chernozhukov2010} to obtain a nondecreasing synthesized quantile function \(\widetilde Q_t\). This step reduces the risk of quantile crossing after independently fitting the quantile-specific models. Evaluating the rearranged quantile function on \(u_r=r/(R+1)\), \(r=1,\ldots,R\), gives the synthesized empirical posterior predictive sample
\[
z_t^{(r)}=\widetilde Q_t(u_r).
\]
The sample \(\{z_t^{(r)}:r=1,\ldots,R\}\) is the per-time marginal predictive distribution used for the displayed synthesis bands, posterior predictive quantiles, and CRPS calculations.

\subsection{Predictive Scoring by CRPS}
\label{subsec:modelselection}

For predictive scoring, we use the Continuous Ranked Probability Score (CRPS) \citep{gneiting2007strictly} as the primary score for forecast-window predictive distributions and, separately, as a preliminary calibration criterion for selected specification choices. The CRPS is negatively oriented, so smaller values are better, and it is a proper score for predictive distributions that rewards both calibration and sharpness.

For a predictive distribution \(G\) with quantile function \(Q_G\) and an observation \(y\), the CRPS is defined as
\[
\text{CRPS}(y, G) = 2 \int_0^1 \rho_u\!\left(y - Q_G(u)\right)\, du,
\qquad
\rho_u(v) = v\big(u - \mathbf{1}\{v<0\}\big),
\]
where \(\rho_u\) is the quantile score, or check loss. Thus, the CRPS is a natural score in the context of quantile estimation. For a generic empirical predictive distribution represented by sample values \(z_t^{(1)},\ldots,z_t^{(S)}\), we approximate the integral by evaluating the empirical quantile function on the grid \(u_s=s/(S+1)\), \(s=1,\ldots,S\). Let \(z_{t,(1)} \leq \cdots \leq z_{t,(S)}\) be the corresponding ordered sample values. Then
\[
\widehat{\mathrm{CRPS}}(y_t,G_t)
= \frac{2}{S}\sum_{s=1}^S
\rho_{u_s}\!\left(y_t-z_{t,(s)}\right).
\]
The same calculation is used for synthesized posterior predictive samples and for raw forecast ensembles, with \(S\) equal to the number of available sample draws or ensemble members at that lead. If the predictive distribution is a point mass at \(x\), the CRPS reduces to \(|y-x|\); averaging this quantity across forecast cases gives the mean absolute error. CRPS is also used in the preliminary pre-cutoff calibration of the transfer-persistence parameter \(\lambda\), as described in Subsection~\ref{subsec:specification}.

\section{FORECASTING THE SAN LORENZO RIVER FLOW}
\label{sec:data}

\subsection{Study Setting and Observations}

\begin{figure}[htbp]
\centering
\includegraphics[width=450pt]{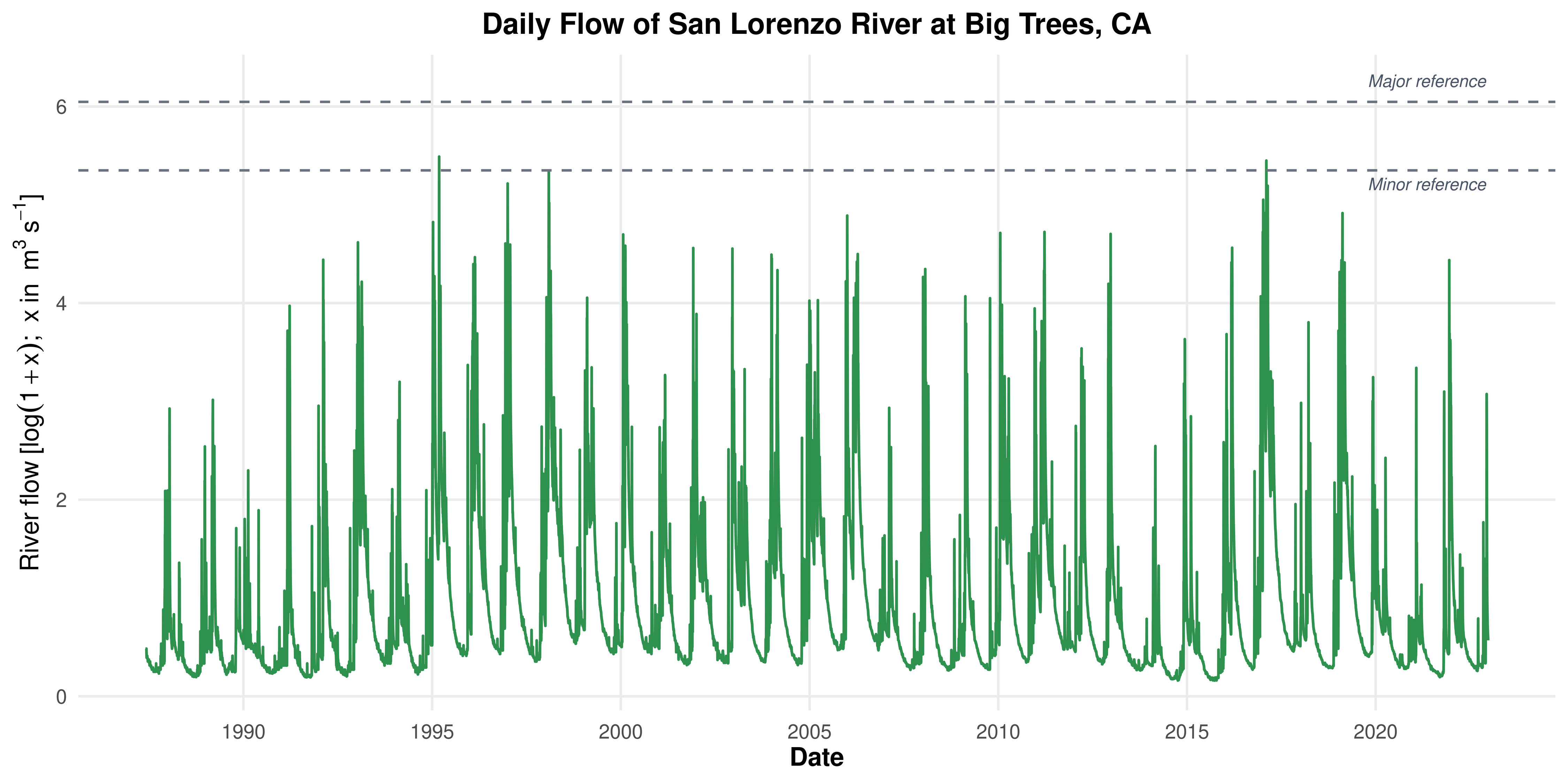}
\caption{
Observed USGS daily water flow on the $\log(1 + x)$ scale, with $x$ measured in m$^3$/s, May 29, 1987 through December 25, 2022. Horizontal dashed lines show approximate current-rating discharge equivalents of the current NWS minor and major stage categories at BTEC1/USGS 11160500, computed from USGS stage-discharge rating 40.0 and transformed to the plotting scale \citep{cnrfc_btec1, usgs_11160500_rating40}. They are operational magnitude references for the discharge scale, not historical flood-stage classifications of daily-mean observations.
}
\label{fig:sanlorenzo}
\end{figure}

Our target series is the average daily natural water flow of the San Lorenzo River at the Big Trees U.S. Geological Survey (USGS) gauging station in Santa Cruz, California (Figure \ref{fig:sanlorenzo}). The raw USGS discharge record is reported in cubic feet per second; we convert it to m$^3$/s before applying the $\log(1+x)$ transformation used in the figures and model inputs. The Big Trees location is characterized by minimal human intervention, so the recorded flow closely reflects natural river dynamics. Over the study window used in this article, the daily series covers May 29, 1987, through December 25, 2022, and contains 12,995 observations.
The San Lorenzo River is a major water resource for the Santa Cruz area and is relevant to both flood response and longer-term water-supply planning. The record includes high-flow episodes as well as prolonged dry conditions. The analysis and forecasting of lower quantiles are therefore relevant for drought conditions, whereas upper quantiles are relevant for flood risk.

We combine the USGS observations with three additional information sources: forecast covariates, retrospective products, and forecast products. Each source plays a different role. The covariates enter the transfer component of the latent process, the retrospective products are used to learn source-specific discrepancies over the pre-cutoff observation window, and the forecast products inform the post-cutoff predictive distributions.

\subsection{External Data Sources and Forecast Covariates}

The transfer component takes three inputs: local precipitation from the PRISM Climate Group \citep{prism}, local soil moisture from ECMWF ERA5-Land \citep{ecmwf}, and a low-dimensional summary of broader climate conditions. Pre-cutoff precipitation and soil moisture are historical gridded covariates extracted at the Big Trees location. After the cutoff, the forecast-window precipitation and soil-moisture covariates are based on Global Ensemble Forecast System (GEFS) precipitation and shallow soil-water forecasts staged in the cutoff-specific origin bundle. These local hydrometeorological covariates enter as deterministic summaries in the present implementation. Precipitation is not modeled through a separate censoring, zero-inflation, or occurrence/intensity layer; dry days are retained in the supplied covariate path, and precipitation contributes through external transfer covariates and deterministic engineered terms. The ERA5-Land soil-moisture series is one of the reanalysis-based model inputs rather than direct observations or uncertainty-free measurements. ERA5/ERA5-Land variables may include short forecast components, but in this analysis they are covariates, not verification observations.

\begin{figure*}[hbt!]
\centering
\centerline{\includegraphics[width=450pt]{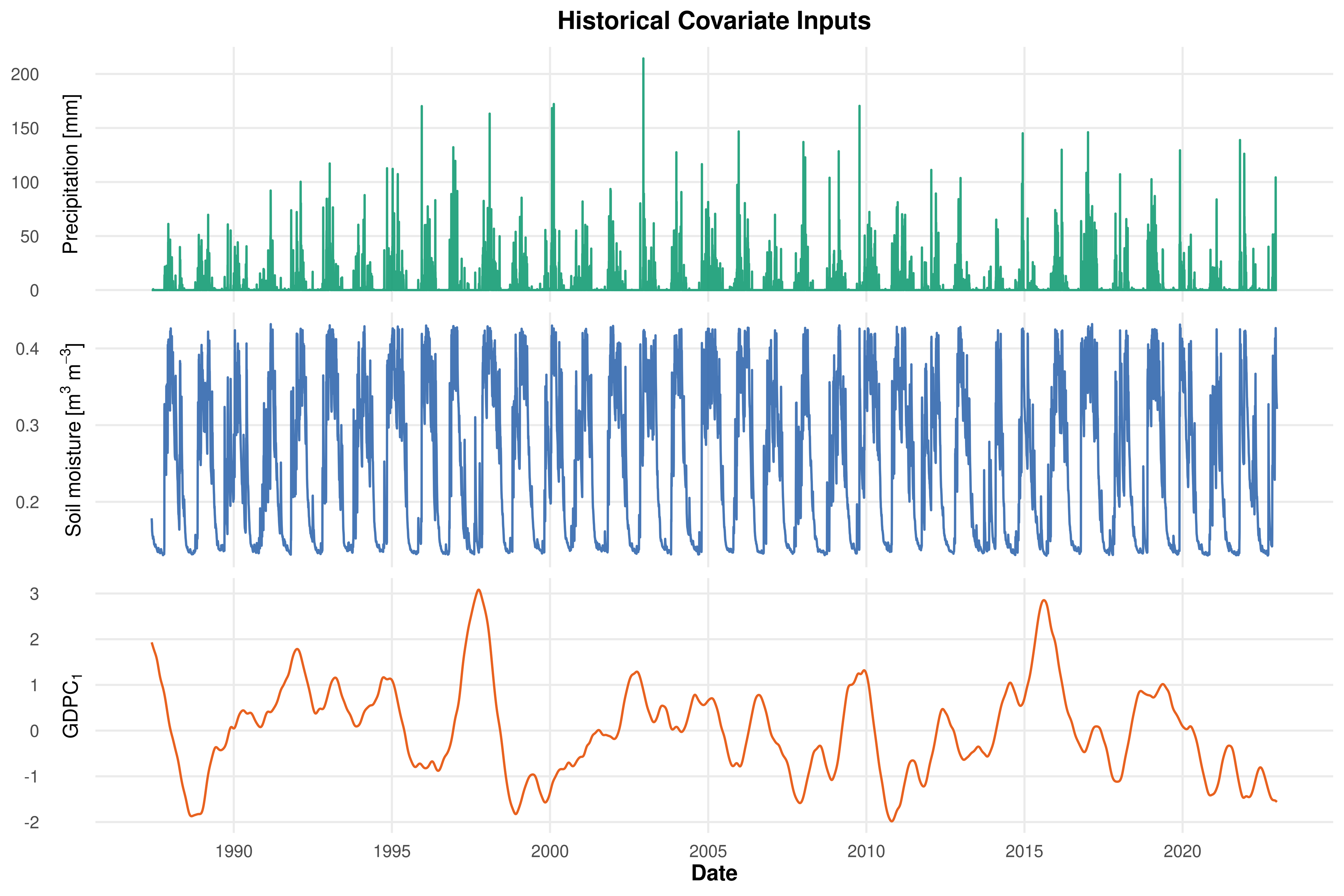}}
\caption{
Historical daily precipitation [mm], soil moisture [m$^3$/m$^3$], and first GDPC score over May 29, 1987 through December 25, 2022.}
\label{fig:covariates}
\end{figure*}

The broader climate summary is constructed from monthly climate and environmental indices from \citet{noaa_climate_indices}, including solar flux, Niño 1+2, the Oceanic Niño Index (ONI), the Southern Oscillation Index (SOI), the Western Hemisphere Warm Pool (WHWP), the Global Mean Temperature (GMT), the Northern Oscillation Index (NOI), the Atlantic Multidecadal Oscillation (AMO), and the Tropical South (TSA) and North Atlantic (TNA) indices. These series are interpolated to daily resolution to align with the hydrological data. We summarize the standardized index matrix with the first generalized dynamic principal component (GDPC) \citep{JSSv092c02}, which explains 44.07\% of the total variability. The GDPC factor is treated as a climate-index covariate, not as an operational forecast product or verification target. Figure~\ref{fig:covariates} shows these three explanatory series.

\begin{figure*}[hbt!]
\centering
\centerline{\includegraphics[width=450pt]{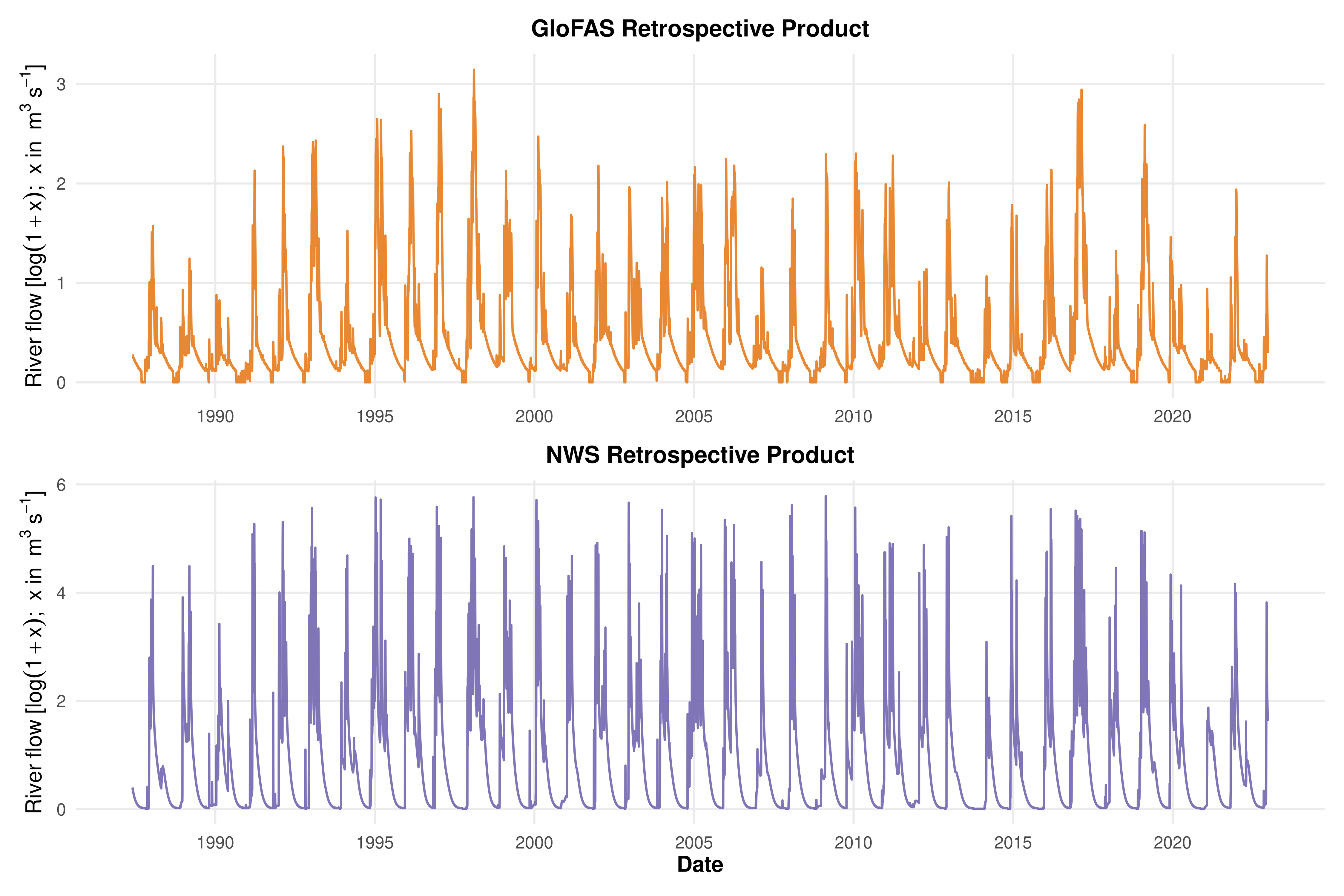}}
\caption{Selected GloFAS and NWS retrospective product series over the available pre-cutoff support window for the December 25, 2022 cutoff, May 29, 1987 through December 25, 2022. Displayed flow values use the $\log(1+x)$ scale with $x$ in m$^3$/s.}
\label{fig:retrospectives}
\end{figure*}

The external product families come from two agencies: GloFAS, operated by ECMWF, and NWS/National Water Model products from NOAA \citep{cosgrove2024noaa}. Each agency contributes both retrospective information and operational forecast information. The retrospective products are useful because they provide historically aligned external series that can be compared against the target observations before the cutoff, allowing the model to learn source-specific discrepancy patterns before those same sources enter the forecast window. The retrospective products enter the pre-cutoff observation window and are used to learn source-specific discrepancies relative to the USGS target series. Figure~\ref{fig:retrospectives} shows the corresponding retrospective series over the available pre-cutoff support window for the December 25, 2022 cutoff.

These products also differ in spatial support. For gridded covariates and GloFAS discharge fields, the workflow extracts the cell or nearest valid river cell associated with the Big Trees gauge coordinates. For NWM streamflow, the workflow uses the corresponding model-output point or reach series and aggregates the latest available forecast issuances to daily member-specific forecast matrices. The PRISM precipitation path uses the 4 km product, while ERA5-Land soil moisture is extracted from its 0.1-degree gridded product. The GloFAS retrospective series are version-specific over the study period, and the reproducible archive records the exact retrospective version and extraction policy used for each cutoff.

The operational forecast products are used only after a cutoff date. The NWS ensemble has 7 members: 6 members provide medium-range forecasts up to 8.5 days, and one control member extends to 10 days. The GloFAS ensemble contains 51 members, all providing medium-range forecasts for up to 28 days. These products therefore differ not only in source but also in forecast horizon and operational update pattern, which is why the forecast window must be modeled source by source rather than as a single homogeneous input stream.

\subsection{Model Specification}
\label{subsec:specification}

For the San Lorenzo application, the latent quantile process combines a local level, three seasonal harmonics, and the transfer component defined in Section~\ref{sec:methodology}. The source-specific discrepancy states use the same trend-and-seasonal backbone. The trend-and-seasonal state has dimension \(p=7\). The matrix \(\mathbf{G}_t\) is composed of a trend component and a seasonal component. For the trend component, we include a constant term to represent the long-term average level of the quantile of interest. This component is defined by the evolution matrix and observational vector:
\( \mathbf{G}^{\text{trend}}_t = [1] \in \mathbb{R}^{1\times1}, \quad
\mathbf{F}^{\text{trend}}_t = [1] \in \mathbb{R}^{1}. \) The seasonal component incorporates three harmonics identified through exploratory spectral analysis to capture various seasonal cycles: an annual harmonic (12 months), a semiannual harmonic (6 months), and a long-term harmonic with an approximately 82-month (6.81-year) cycle that captures cyclical behavior associated with alternating phases of heavy rainfall and prolonged drought. The seasonal component's evolution matrix and observational vector are defined as:
\( \mathbf{G}^{\text{seas}}_t = \text{BlockDiag}\left(\textbf{J}_2(1,1),\ \textbf{J}_2(1,2),\ \textbf{J}_2\left(1,\frac{1}{6.8068493}\right)\right),\quad
\mathbf{F}^{\text{seas}}_t  =
\begin{bmatrix}
\mathbf{E}_2 \\[5pt]
\mathbf{E}_2 \\[5pt]
\mathbf{E}_2
\end{bmatrix}, \)
where
\( \textbf{J}_2(1,\omega) =
\begin{bmatrix}
\cos(\omega) & \sin(\omega) \\[5pt]
-\sin(\omega) & \cos(\omega)
\end{bmatrix},\quad
\mathbf{E}_2 =
\begin{bmatrix}
1 \\[5pt]
0
\end{bmatrix}. \)
Here \(\textbf{J}_2(1,\omega) \in \mathbb{R}^{2\times2}\), \(\mathbf{G}^{\text{seas}}_t \in \mathbb{R}^{6\times6}\), \(\mathbf{F}^{\text{seas}}_t \in \mathbb{R}^{6}\), and \(\mathbf{E}_2 \in \mathbb{R}^{2}\). Combining both components, the complete evolution matrix and observational vector for trend and seasonal dynamics become:
\(
\mathbf{G}_t = \text{BlockDiag}\left(\mathbf{G}^{\text{trend}}_t,\ \mathbf{G}^{\text{seas}}_t\right),\quad
\mathbf{F}_t =
\begin{bmatrix}
\mathbf{F}^{\text{trend}}_t \\[5pt]
\mathbf{F}^{\text{seas}}_t
\end{bmatrix}.
\) Thus, \(\mathbf{G}_t \in \mathbb{R}^{7\times7}\) and \(\mathbf{F}_t \in \mathbb{R}^{7}\).

For prior hyperparameters (see Subsection~\ref{subsec:prioranddisc}), we specify \(\phi^s = 10^6\), \(\nu^s = 1\), and \(a_{\sigma^s} = b_{\sigma^s} = 10^{-6}\) for all \(s\in\mathcal{J}_0\), reflecting diffuse prior beliefs for both the USGS observation channel and the external product channels. Sensitivity analyses for \(\phi^s \in \{10^1, 10^3, 10^6\}\) and \(b_{\sigma^s} \in \{10^{-1}, 10^{-3}, 10^{-6}\}\) showed no noticeable impact on model fit.

We choose discount factors componentwise to reflect each component's time scale and role. Recall that, in the discount-factor formulation \citep{prado2021time}, a smaller value \(d \in (0,1]\) for a given state parameter implies larger state-evolution variance and therefore greater adaptability of that component. The selected specifications use high, near-one discount factors for the trend, seasonal, discrepancy, transfer, and covariate-effect states. These values preserve persistent long-run behavior while allowing slow adaptation across the multi-decade observation window.

The transfer-persistence parameter \(\lambda\) was selected in a preliminary one-dimensional calibration step for the selected exAL-M-T1 specification. To avoid confounding transfer persistence with state-evolution adaptability, the component discount factors were held fixed during this step. We considered the grid \(\lambda \in \{0.60,0.61,\ldots,0.99\}\) and compared candidate values using CRPS for the median-quantile model over the pre-cutoff observational window at the representative December 25, 2022 cutoff, which has the longest pre-origin archive among the rolling-origin cases. The selected value was \(\lambda=0.97\), which was then held fixed across the reported quantile levels and forecast origins. This calibration is a model-specification step rather than a nested tuning procedure repeated inside each forecast-origin evaluation; the forecast-window CRPS values reported below evaluate the fixed calibrated specification against held-out future USGS observations.

For the forecast-window covariance prior, the parameter \(\epsilon\) controls the prior weight assigned to the covariance structure learned before the cutoff. Smaller values give a weaker anchor and allow greater adaptation during forecast propagation, while larger values retain more of the pre-cutoff covariance structure. For the representative December 25, 2022 selected specification, we set \(\epsilon=1\) and \(c=1\). These settings stabilize short-horizon forecast propagation while allowing the forecast-window state evolution to adapt during a winter origin with substantial rainfall-driven uncertainty.

\subsection{Rolling-Origin Forecast Evaluation Design}

The forecast evaluation uses five cutoff-specific, version-consistent staged datasets that span contrasting hydrological conditions and remain compatible with the archived retrospective and forecast products available from each agency. The retained origins include relatively low-flow windows as well as winter high-flow episodes, so the assessment covers different operating regimes, although it is not a continuous daily hindcast over the full post-2022 period. Each cutoff fixes a forecast origin, uses only information available at that origin to fit seven quantile-specific models and synthesize the corresponding posterior predictive distribution, and scores that distribution against future USGS observations held out over the forecast window.

Constructing one staged dataset requires more than shifting the cutoff date in a fixed table. The archived products come from evolving operational systems with different release histories, product versions, update frequencies, horizons, ensemble-member structures, spatial supports, and access interfaces. For each retained origin we therefore rebuild a version-consistent bundle containing USGS observations, retrospective products, issued forecast products, and forecast-window covariates. This archive-reconstruction step is the main practical constraint on a denser rolling-origin design: many more origins would require substantially more data recovery, version matching, spatial extraction, covariate staging, and computation, while nearby forecast windows would repeatedly evaluate similar hydrological episodes. We therefore use archive-feasible, version-consistent origins that span contrasting hydrological settings and avoid dense overlaps that would overrepresent the same episode.

At a cutoff \(c\), the model is fit using USGS observations and retrospective products available through \(c\). The forecast-window predictive distributions are then synthesized using the latest forecast products issued at or before \(c\), the forecast-window precipitation and soil-moisture covariates from the same staged origin bundle, and the canonical GDPC climate-index factor sliced consistently to that bundle. The local forecast-window covariates are GEFS ensemble-based precipitation and shallow soil-water covariates reduced to deterministic summaries before entering the transfer component. The GDPC factor is not treated as an operational forecast product or verification target. Post-cutoff USGS observations are reserved strictly for verification and are not used to fit or update the predictive distributions.

GloFAS issues forecast products daily, whereas NWS issues them hourly. For GloFAS, the forecast matrix is the daily issue associated with the cutoff. For NWS, we first restrict to issuances available by the cutoff, retain the most recent available issuance for each target time and ensemble member, and then aggregate the retained hourly values to daily resolution. Thus older forecast issuances are not averaged into the publication forecast matrices; any older retained labels are kept only for workflow file-name compatibility. Figure~\ref{fig:ensembles} illustrates the resulting information structure for the representative cutoff on December 25, 2022 in a compact forecast-context display that spans 28 days before and 28 days after the cutoff. The GloFAS forecast window extends from December 26, 2022 to January 22, 2023, whereas the NWS forecast window extends from December 26, 2022 to January 2, 2023.
\section{FORECAST VALIDATION RESULTS}
\label{sec:forecastvalidation}

\subsection{Comparative Forecast Performance}
\label{subsec:benchmarkresults}

For each of the five cutoffs, we evaluate forecasts over two horizons: eight days and 28 days. The eight-day horizon is the common daily horizon available for the NWS product, whereas the 28-day horizon matches the GloFAS medium-range forecast window used in the main comparison. For each case we consider nine Bayesian variants of the common state-space framework. These variants differ only in three respects: the observation likelihood, whether the synthesis uses only the USGS observation channel or all source channels jointly, and whether the transfer component remains active during the forecast window. We summarize them with the label \(L\)-\(S\)-\(T\), where \(L\in\{\mathrm{N},\mathrm{AL},\mathrm{exAL}\}\) denotes a Gaussian, asymmetric Laplace, or extended asymmetric Laplace observation likelihood, \(S\in\{\mathrm{U},\mathrm{M}\}\) indicates whether the synthesis is univariate or multivariate, and \(T\in\{\mathrm{T0},\mathrm{T1}\}\) indicates whether the transfer component is suppressed or retained during the forecast window.

For the asymmetric Laplace and extended asymmetric Laplace variants, we fit the framework separately at seven target quantile levels: \(0.05\), \(0.20\), \(0.35\), \(0.50\), \(0.65\), \(0.80\), and \(0.95\). The Gaussian variants correspond to normal dynamic linear models. Forecast skill is evaluated from the resulting posterior predictive distributions by the mean continuous ranked probability score (CRPS). In all cases we use the variational Bayes procedure described in Subsection~\ref{subsec:postinf}.

Table~\ref{tab:benchmark_crps_models} reports mean 28-day forecast-window CRPS. The multivariate rows use all available source channels, including NWS forecast information over the leads for which it is available, and the AL/exAL rows are posterior predictive syntheses assembled from the seven fitted quantile-specific models. The raw GloFAS product is included as the horizon-compatible 28-day raw reference; raw NWS is handled in the separate eight-day comparison because the archived NWS forecasts provide only eight valid daily leads for these origins.

\begin{table*}[htbp]
\centering
\renewcommand{\arraystretch}{1.08}
\begin{threeparttable}
\caption{Mean 28-day forecast-window CRPS by model family and cutoff across the five rolling-origin evaluation folds, including normal dynamic linear model baselines. Lower values are better; bold indicates the lowest CRPS within each cutoff column.}
\label{tab:benchmark_crps_models}
\begin{tabular*}{\textwidth}{@{\extracolsep{\fill}} >{\ttfamily}l r r r r r}
\toprule
Model label & 01/23/2021 & 11/12/2021 & 12/21/2021 & 05/11/2022 & 12/25/2022 \\
\midrule
\multicolumn{6}{l}{\textit{Raw forecast products}} \\
RAW-GLOFAS & 0.40366 & 0.16957 & 0.68246 & 0.27227 & 1.56006 \\
\midrule
\multicolumn{6}{l}{\textit{Bayesian benchmark variants}} \\
N-U-T1 & 0.33592 & 0.17059 & 1.19344 & 0.15077 & 2.49968 \\
N-M-T0 & 1.84333 & 0.38023 & 0.65964 & 0.67014 & 0.64404 \\
N-M-T1 & 3.21490 & 0.89096 & 3.04362 & 0.86820 & 3.88863 \\
AL-U-T1 & 0.23107 & 0.07795 & 0.87049 & 0.10506 & 1.73376 \\
AL-M-T0 & 0.46798 & 0.19990 & 0.58668 & 0.09940 & 1.33703 \\
AL-M-T1 & 0.14592 & 0.05551 & 0.27775 & 0.05467 & 0.62764 \\
exAL-U-T1 & 0.21622 & 0.10194 & 0.83337 & 0.12452 & 1.70548 \\
exAL-M-T0 & 0.75682 & 1.72135 & 0.97762 & 1.02087 & 1.21132 \\
exAL-M-T1 & \textbf{0.13971} & \textbf{0.04724} & \textbf{0.26045} & \textbf{0.02273} & \textbf{0.53806} \\
\bottomrule
\end{tabular*}
\end{threeparttable}
\end{table*}

Table~\ref{tab:benchmark_crps_models} shows a consistent pattern across the five cutoffs. The \(U\) versus \(M\) comparison isolates the contribution of external source channels beyond the USGS observation channel alone, whereas the \(T0\) versus \(T1\) comparison isolates the value of retaining the transfer component during the forecast window. The two multivariate models with an active forecast-window transfer component, AL-M-T1 and exAL-M-T1, provide the strongest performance in the 28-day synthesis comparison. The exAL-M-T1 specification attains the lowest 28-day CRPS at all five rolling-origin cutoffs, and we use it as the reference specification for the remaining model summaries.

\begin{table*}[htbp]
\centering
\renewcommand{\arraystretch}{1.08}
\begin{threeparttable}
\caption{Mean CRPS over the common eight-day NWS forecast horizon. Lower values are better; bold indicates the lowest CRPS within each cutoff column.}
\label{tab:benchmark_crps_models_nws_horizon}
\begin{tabular*}{\textwidth}{@{\extracolsep{\fill}} >{\ttfamily}l r r r r r}
\toprule
Model label & 01/23/2021 & 11/12/2021 & 12/21/2021 & 05/11/2022 & 12/25/2022 \\
\midrule
\multicolumn{6}{l}{\textit{Raw forecast products}} \\
RAW-GLOFAS & 0.57251 & 0.20670 & 1.29127 & 0.32274 & 1.48188 \\
RAW-NWS & 0.83037 & 1.37192 & \textbf{0.28118} & 0.28366 & \textbf{0.55680} \\
\midrule
\multicolumn{6}{l}{\textit{Bayesian benchmark variants}} \\
N-U-T1 & 0.76838 & 0.13732 & 2.04040 & 0.12611 & 1.87805 \\
N-M-T0 & 1.31032 & 0.25581 & 0.41461 & 0.63732 & 1.00096 \\
N-M-T1 & 4.72798 & 0.71930 & 5.13027 & 0.79737 & 4.53177 \\
AL-U-T1 & \textbf{0.18835} & 0.10367 & 0.75625 & 0.03599 & 0.62002 \\
AL-M-T0 & 0.63929 & 0.16893 & 1.15588 & 0.09506 & 1.00922 \\
AL-M-T1 & 0.23740 & 0.07170 & 0.56351 & 0.05867 & 0.61593 \\
exAL-U-T1 & 0.18999 & 0.11312 & 0.77939 & 0.04489 & 0.62982 \\
exAL-M-T0 & 0.94687 & 1.69464 & 0.69022 & 1.02364 & 1.31568 \\
exAL-M-T1 & 0.22691 & \textbf{0.05858} & 0.59445 & \textbf{0.02567} & 0.59070 \\
\bottomrule
\end{tabular*}
\begin{tablenotes}
\item \textit{Note:} This table restricts every row to forecast leads 1--8, the common daily horizon available for NWS, GloFAS, and the Bayesian predictive distributions. It is therefore the appropriate direct comparison to \texttt{RAW-NWS}; Table~\ref{tab:benchmark_crps_models} gives the complementary 28-day comparison and omits NWS.
\end{tablenotes}
\end{threeparttable}
\end{table*}

The short-horizon comparison in Table~\ref{tab:benchmark_crps_models_nws_horizon} is the appropriate direct comparison to the raw NWS product. It preserves the same held-out USGS verification target and transformed scale as the 28-day table, but restricts all rows to the NWS-available forecast horizon. The comparison is more mixed than the 28-day table: raw NWS is strongest at the December 2021 and December 2022 cutoffs, AL-U-T1 is strongest in January 2021, and exAL-M-T1 is strongest in November 2021 and May 2022. This confirms that the NWS product can be competitive over its shorter native horizon while the full 28-day synthesis comparison remains a distinct evaluation problem.

Table~\ref{tab:he4_quantile_check_loss} complements the CRPS comparisons with targeted quantile diagnostics based on check loss at the fitted quantile levels. We consider the four principal quantile-synthesis competitors: exAL-M-T1, AL-M-T1, exAL-U-T1, and AL-U-T1. These diagnostics use the same held-out USGS forecast-window observations and the same transformed discharge scale as Table~\ref{tab:benchmark_crps_models}. The multivariate synthesis models have lower check loss than the univariate references in most cutoff--quantile cells. The exAL-M-T1 row is strongest in most cells, while AL-M-T1 remains competitive in selected upper-tail cases.

\begin{table*}[htbp]
\centering
\renewcommand{\arraystretch}{1.08}
\begin{threeparttable}
\caption{Mean forecast-window quantile check loss by synthesis model, cutoff, and target quantile. Lower values are better; bold indicates the lowest check loss within each cutoff and quantile column.}
\label{tab:he4_quantile_check_loss}
\begin{tabular*}{\textwidth}{@{\extracolsep{\fill}} >{\ttfamily}l r r r r r r r}
\toprule
Model & q0.05 & q0.20 & q0.35 & q0.50 & q0.65 & q0.80 & q0.95 \\
\midrule
\multicolumn{8}{l}{\textit{Cutoff 01/23/2021}} \\
exAL-M-T1 & \textbf{0.02529} & \textbf{0.07061} & \textbf{0.09149} & \textbf{0.09407} & \textbf{0.08697} & 0.07165 & \textbf{0.03093} \\
AL-M-T1 & 0.02894 & 0.07178 & 0.09341 & 0.09700 & 0.08931 & \textbf{0.06937} & 0.04116 \\
exAL-U-T1 & 0.03314 & 0.09434 & 0.12252 & 0.13140 & 0.13265 & 0.12867 & 0.10144 \\
AL-U-T1 & 0.03150 & 0.08874 & 0.12511 & 0.13534 & 0.14456 & 0.14617 & 0.12576 \\
\addlinespace[1pt]
\multicolumn{8}{l}{\textit{Cutoff 11/12/2021}} \\
exAL-M-T1 & \textbf{0.00943} & 0.02226 & \textbf{0.02164} & \textbf{0.03224} & \textbf{0.03659} & \textbf{0.02737} & 0.01267 \\
AL-M-T1 & 0.01114 & 0.02242 & 0.02264 & 0.03547 & 0.03937 & 0.03394 & 0.02369 \\
exAL-U-T1 & 0.01110 & 0.02586 & 0.08966 & 0.09603 & 0.06947 & 0.04183 & 0.01279 \\
AL-U-T1 & 0.01843 & \textbf{0.02075} & 0.07354 & 0.06746 & 0.04860 & 0.02874 & \textbf{0.01018} \\
\addlinespace[1pt]
\multicolumn{8}{l}{\textit{Cutoff 12/21/2021}} \\
exAL-M-T1 & \textbf{0.02204} & \textbf{0.08243} & \textbf{0.13306} & \textbf{0.17423} & \textbf{0.21136} & 0.17833 & 0.08278 \\
AL-M-T1 & 0.05594 & 0.10246 & 0.14994 & 0.17786 & 0.22531 & \textbf{0.17118} & \textbf{0.05205} \\
exAL-U-T1 & 0.06544 & 0.22311 & 0.33019 & 0.42750 & 0.54414 & 0.64406 & 0.67338 \\
AL-U-T1 & 0.06384 & 0.23665 & 0.33938 & 0.45254 & 0.57593 & 0.67517 & 0.68584 \\
\addlinespace[1pt]
\multicolumn{8}{l}{\textit{Cutoff 05/11/2022}} \\
exAL-M-T1 & \textbf{0.00754} & \textbf{0.00818} & \textbf{0.00784} & \textbf{0.01526} & \textbf{0.01387} & \textbf{0.01399} & \textbf{0.00904} \\
AL-M-T1 & 0.01095 & 0.03552 & 0.03957 & 0.03433 & 0.02135 & 0.02730 & 0.01661 \\
exAL-U-T1 & 0.01852 & 0.04289 & 0.10852 & 0.10902 & 0.07885 & 0.04803 & 0.01501 \\
AL-U-T1 & 0.03186 & 0.03619 & 0.09686 & 0.08487 & 0.06068 & 0.03675 & 0.01363 \\
\addlinespace[1pt]
\multicolumn{8}{l}{\textit{Cutoff 12/25/2022}} \\
exAL-M-T1 & \textbf{0.08853} & \textbf{0.21289} & \textbf{0.30922} & \textbf{0.41184} & 0.41426 & 0.25746 & 0.14369 \\
AL-M-T1 & 0.10685 & 0.33826 & 0.49367 & 0.55801 & \textbf{0.28439} & \textbf{0.22049} & \textbf{0.13795} \\
exAL-U-T1 & 0.12106 & 0.41104 & 0.63858 & 0.86423 & 1.10845 & 1.33342 & 1.47893 \\
AL-U-T1 & 0.12012 & 0.43475 & 0.64594 & 0.88340 & 1.13232 & 1.35815 & 1.46909 \\
\bottomrule
\end{tabular*}
\begin{tablenotes}
\item \textit{Note:} Check loss is computed on forecast-window rows only, using held-out USGS observations as verification targets on the same $\log(1+Q)$ scale used for CRPS. The four synthesis competitors are resolved directly from the frozen HE-2 publication manifest.
\end{tablenotes}
\end{threeparttable}
\end{table*}

Appendix~\ref{app:he3ablation} reports a component-removal sensitivity analysis of the selected extended-likelihood multivariate model. The analysis preserves the same rolling-origin cutoffs, input construction, likelihood, and inference scheme while removing one structural component at a time. Its role is illustrative, and the benchmark tables above remain the primary forecast-comparison evidence; comparative forecast evaluation remains the main empirical evidence.

\section{INTERPRETATION OF THE SELECTED SPECIFICATION}
\label{sec:interpretation}

We focus on exAL-M-T1 because it has the lowest 28-day forecast-window CRPS across the five rolling-origin cutoffs in Table~\ref{tab:benchmark_crps_models}.

Detailed source-specific summaries for the exAL shape and scale parameters are reported in Appendix~\ref{app:sourceparams}. In the main text, we focus on the most interpretable features of the selected model.

\subsection{Covariate Effects}

\begin{table*}[htbp]
\centering
\begin{threeparttable}
\caption{Selected Posterior Means and 95\% Credible Intervals for Transfer-Function Covariates}
\label{tab:components_23_31}
\begin{tabular*}{\textwidth}{@{\extracolsep{\fill}} l c S[table-format=-1.5] l }
\toprule
\textbf{Covariate} & \textbf{Quantile} & \textbf{Mean} & \textbf{95\% CI} \\
\midrule
\multirow{3}{*}{Precipitation}  & 5th & -0.10282 & $[-0.10724,\ -0.09853]$ \\
& 50th & -0.06258 & $[-0.06830,\ -0.05698]$ \\
& 95th & 0.08632 & $[0.08044,\ 0.09208]$ \\
\midrule
\multirow{3}{*}{Soil Moisture}  & 5th & -0.02695 & $[-0.03315,\ -0.02082]$ \\
& 50th & -0.03325 & $[-0.04149,\ -0.02512]$ \\
& 95th & 0.01462 & $[0.00546,\ 0.02384]$ \\
\midrule
\multirow{3}{*}{First GDPC factor}  & 5th & -0.00198 & $[-0.00202,\ -0.00193]$ \\
& 50th & -0.00132 & $[-0.00137,\ -0.00126]$ \\
& 95th & -0.00116 & $[-0.00122,\ -0.00110]$ \\
\bottomrule
\end{tabular*}
\begin{tablenotes}
\item \textit{Note:} Posterior means and 95\% credible intervals $[{\rm Q2.5}, {\rm Q97.5}]$ for selected transfer-function coefficients at the representative December 25, 2022 cutoff, for the 5th, 50th, and 95th quantile models. \texttt{First GDPC factor} refers to the first generalized dynamic principal component introduced in Section~\ref{sec:data}.
\end{tablenotes}
\end{threeparttable}
\end{table*}

Table~\ref{tab:components_23_31} reports posterior means and 95\% credible intervals for the three transfer-function covariate coefficients displayed in the main text at the 5th, 50th, and 95th quantiles. For the representative December 25, 2022 cutoff, the local covariates do not act as a common upward shift across the fitted quantile levels. Precipitation has negative coefficients at the 5th and 50th quantiles and a positive coefficient at the 95th quantile; soil moisture shows the same sign pattern, with negative coefficients at the 5th and 50th quantiles and a positive coefficient at the 95th quantile.
These estimates are therefore best interpreted as quantile-specific transfer effects. In this representative fit, local precipitation and soil moisture modify the shape of the fitted conditional quantile process, with positive upper-tail coefficients but negative lower- and central-quantile coefficients on the transformed scale. The first GDPC factor is negative at the displayed quantiles and smaller on the reported coefficient scale, so we do not interpret it as the primary source of the displayed upper-tail transfer behavior.

\subsection{Selected-Model Dynamics Across Representative Regimes}

To illustrate how the model behaves across contrasting hydrological conditions, we show posterior quantile dynamics for the 2022-12-25 cutoff exAL-M-T1 specification. The displayed windows cover the 2012--2016 drought and the 2017--2019 wet period. These figures are interpretation diagnostics: they show how the selected model decomposes fitted historical quantile behavior across contrasting regimes.

\begin{figure}[htbp]
\centering
\includegraphics[width=450pt]{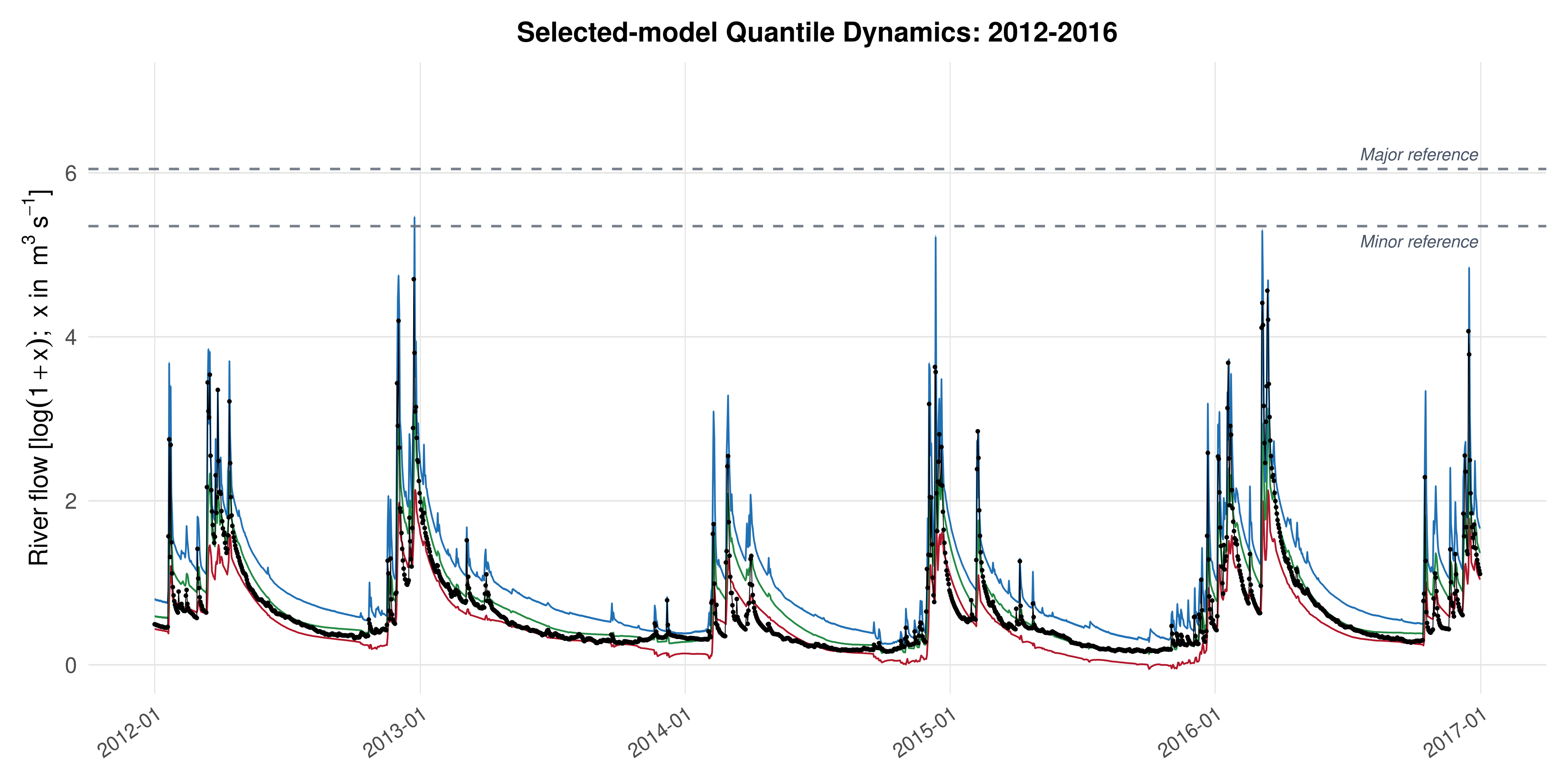}
\caption{
Posterior mean and 95\% credible intervals of river-flow quantiles during the dry period (2012--2016), estimated from the 2022-12-25 cutoff exAL-M-T1 model. Shown are the 5th (red), 50th (green), and 95th (blue) quantiles together with observed daily USGS flow (black). Horizontal dashed lines show the same current-rating NWS minor and major discharge-reference levels as in Figure~\ref{fig:sanlorenzo}.
}
\label{fig:dry_quantile}
\end{figure}

\begin{figure}[htbp]
\centering
\includegraphics[width=450pt]{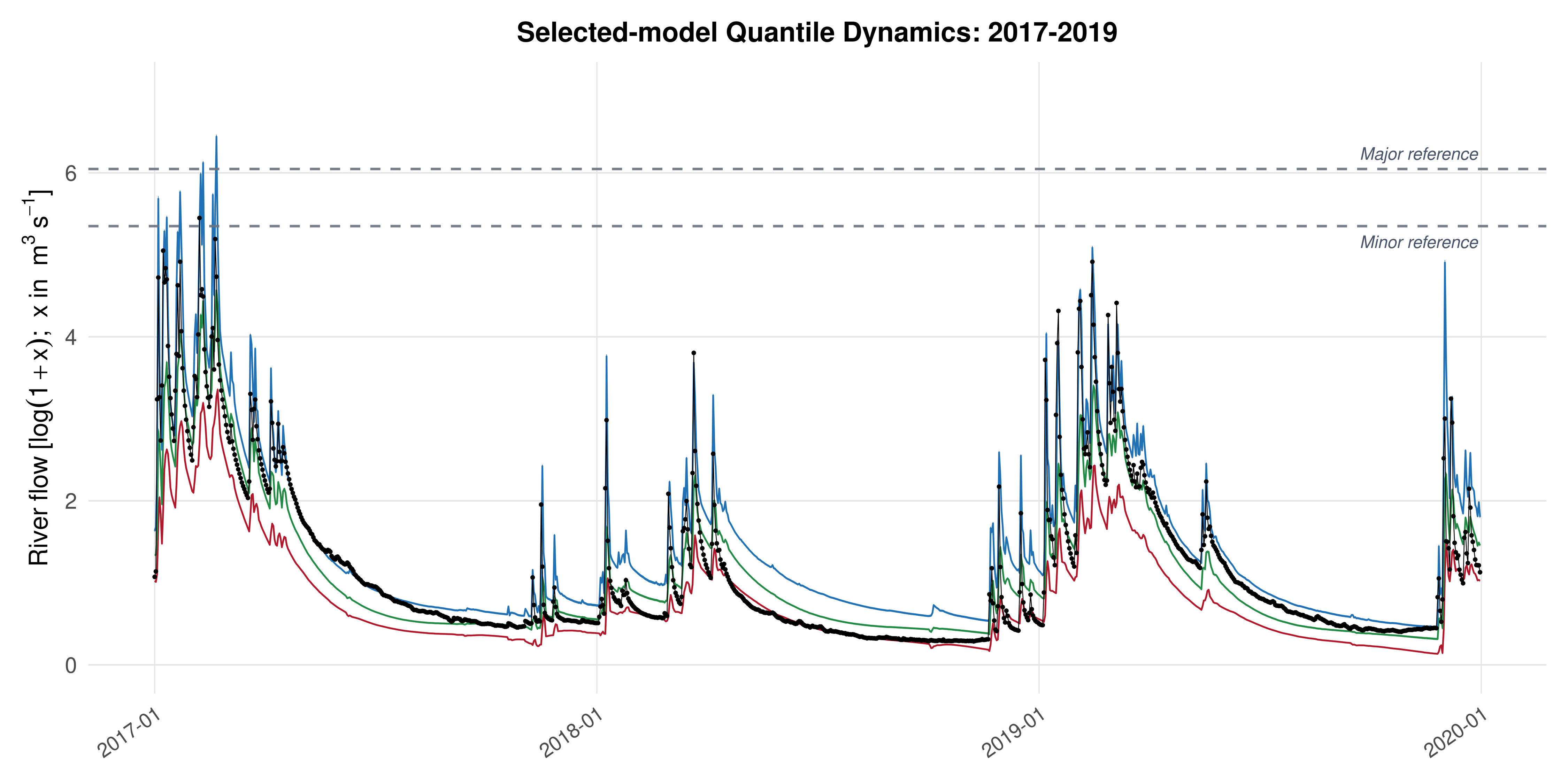}
\caption{
Posterior mean and 95\% credible intervals of river-flow quantiles during the wet period (2017--2019), estimated from the 2022-12-25 cutoff exAL-M-T1 model. Shown are the 5th (red), 50th (green), and 95th (blue) quantiles together with observed daily USGS flow (black). Horizontal dashed lines show the same current-rating NWS minor and major discharge-reference levels as in Figure~\ref{fig:sanlorenzo}.
}
\label{fig:rainy_quantile}
\end{figure}

Figures~\ref{fig:dry_quantile} and \ref{fig:rainy_quantile} compare the fitted posterior quantile dynamics during the drought and wet periods. During 2012--2016, the posterior bands remain comparatively narrow and the upper fitted quantile rarely approaches the current-rating NWS discharge-reference levels. During 2017--2019, winter peaks produce wider separation between the lower and upper fitted quantile curves, and the 95th quantile reaches levels comparable to the current-rating major-stage discharge reference in 2017 and 2019. This contrast indicates stronger upper-tail and cross-quantile variation during wetter regimes.

\begin{figure}[htbp]
\centering
\includegraphics[width=450pt]{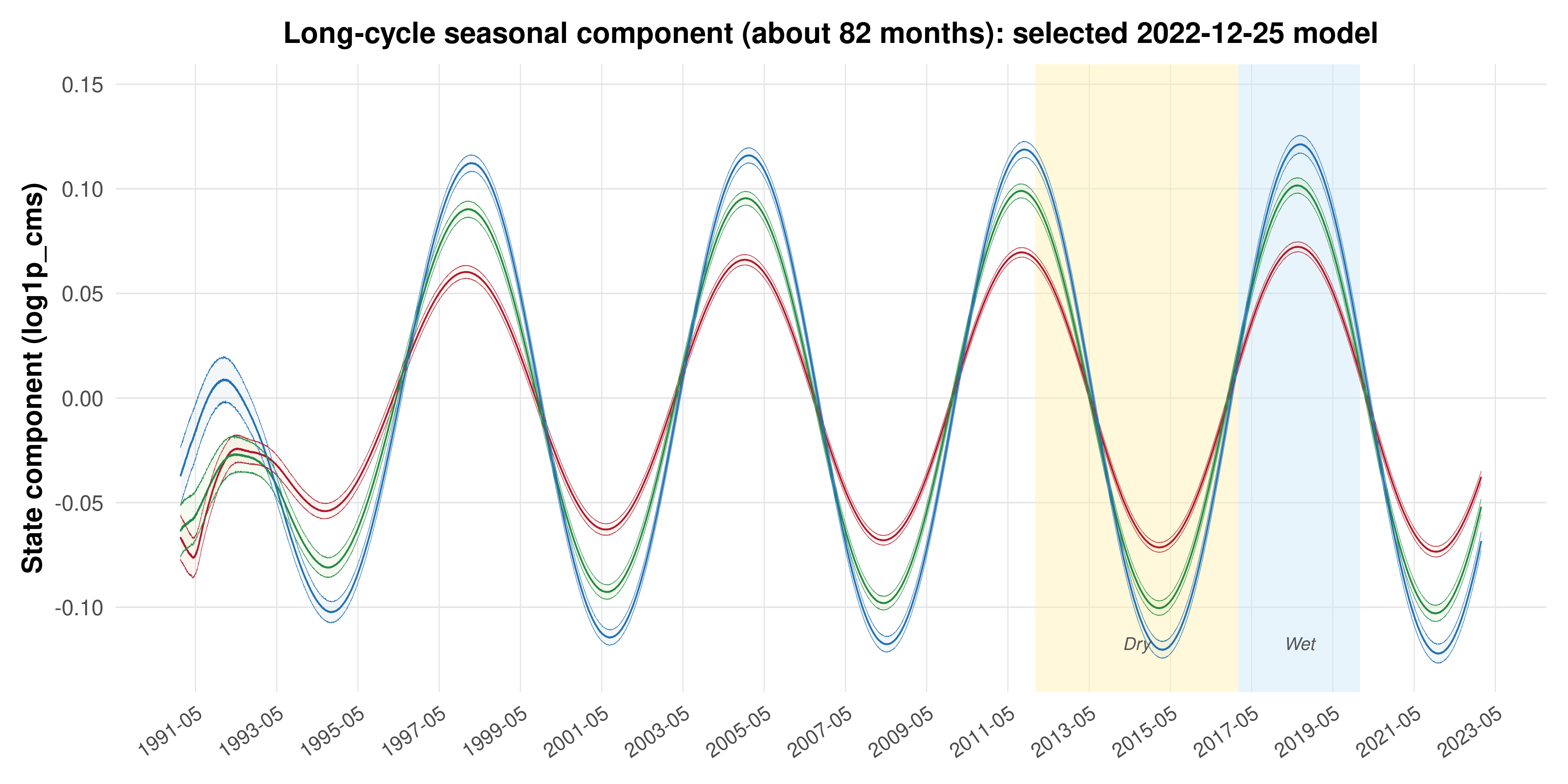}
\caption{
Selected-model posterior mean and 95\% credible intervals for the long-cycle seasonal component, with period approximately 6.81 years, or about 82 months, estimated from the 2022-12-25 cutoff exAL-M-T1 model. Estimates are shown for the 5th (red), 50th (green), and 95th (blue) quantiles. Shaded regions mark the 2012--2016 dry period and the 2017--2019 wet period.
}
\label{fig:80_components}
\end{figure}

A complementary diagnostic is the long-cycle seasonal state. Figure~\ref{fig:80_components} shows this component alone for the same three representative quantile levels. The retained noninteger seasonal frequency corresponds to a period of approximately 6.81 years, or about 82 months. The component is broadly aligned across the three quantiles and tracks multi-year dry and wet phases in the historical record, illustrating how the selected state-space model separates long-cycle behavior from trend, transfer effects, and source-specific corrections.

\clearpage

\subsection{Predictive Synthesis Illustration}
\label{subsec:synthillustration}

\begin{figure}[htbp]
\centering
\includegraphics[width=450pt]{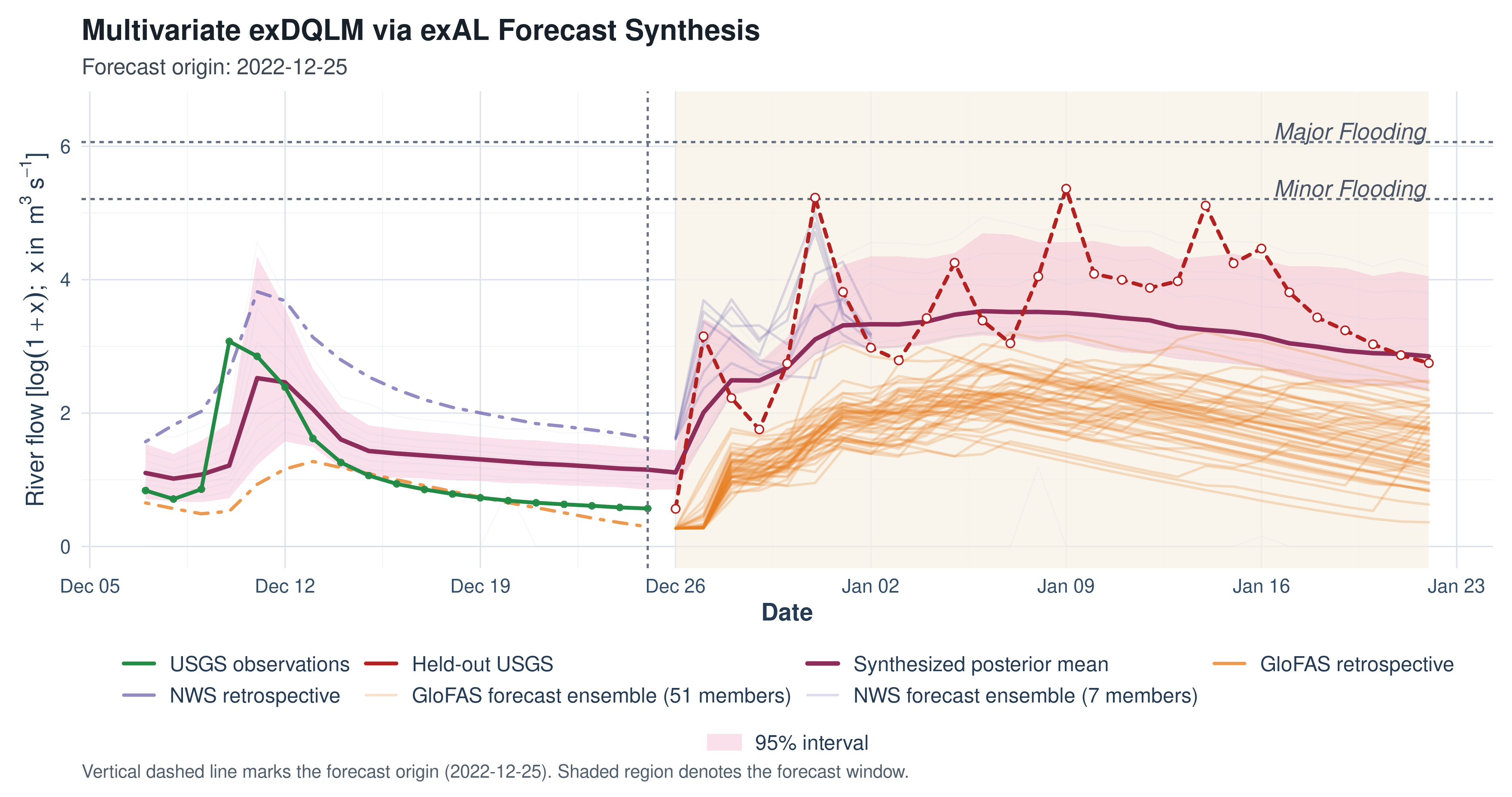}
\caption{
Synthesized posterior predictive distribution for river flow on the $\log(1+x)$ scale, with $x$ measured in m$^3$/s, at the December 25, 2022 forecast origin under the exAL-M-T1 model. Observed USGS measurements before the origin are shown in green, and held-out USGS verification after the origin is shown in red. The pink band is the central 95\% posterior predictive credible band from the synthesized distribution, and the dark center line summarizes the corresponding synthesized center. Thin overlaid lines correspond to displayed forecast quantiles at levels 0.05, 0.20, 0.35, 0.50, 0.65, 0.80, and 0.95. The raw retrospective and forecast products used at this cutoff are shown for context.
}
\label{fig:synth1}
\end{figure}

The posterior bands in Figures~\ref{fig:dry_quantile} and \ref{fig:rainy_quantile} summarize uncertainty around fitted quantile-location curves. Figure~\ref{fig:synth1} instead shows the synthesized posterior predictive distribution for the December 25, 2022 cutoff, after combining the quantile-specific forecasts, source corrections, and forecast-window inputs. It illustrates how the fitted quantile-specific forecasts combine into one predictive distribution for a given forecast origin.

Because exAL-M-T1 is the selected extended-likelihood multivariate specification, Figure~\ref{fig:synth1} is the main predictive-synthesis illustration. The synthesized predictive distribution combines observed USGS measurements before the cutoff with the retrospective and forecast products available at that origin. The upper quantiles rise sharply after December 25, 2022 and reach levels comparable to the current-rating major-stage discharge reference, reflecting the high-flow risk in this forecast window. Unlike the fitted historical quantile-location summaries, the posterior predictive envelope can vary across the forecast window as the source-specific corrections and forecast-window covariates change. A univariate transfer-active reference is reported in Appendix~\ref{app:historicalsynth} to separate the multivariate operational synthesis from a simpler reference that excludes retrospective-product and forecast-product source channels.

\section{Conclusions}
\label{sec5}

We developed a Bayesian quantile-based correction-and-synthesis framework for river-flow forecasting that links USGS observations, retrospective products, and forecast products through a shared latent quantile process. The framework combines source-specific discrepancy learning over the pre-cutoff observation window with a forecast-window synthesis step that propagates discrepancy corrections learned from retrospective products together with forecast-window transfer covariates. Variational Bayes based on Laplace--Delta approximations provides a scalable alternative to full MCMC for fitting the extended dynamic quantile linear model.

In the San Lorenzo River application, the main empirical result is the five-cutoff rolling-origin forecast comparison based on the latest forecast products available at each cutoff. Relative to simpler Bayesian variants, the multivariate specifications with an active forecast-window transfer component provide the strongest performance in the 28-day synthesis comparison. The selected exAL-M-T1 specification attains the lowest 28-day forecast-window CRPS in all five cutoffs. A separate eight-day NWS-horizon table preserves the direct operational comparison to NWS without mixing forecast horizons. Taken together, the results support the value of source-aware multivariate correction for the 28-day forecast window, while the shorter NWS-horizon comparison shows that gains are not uniform across every origin and horizon.

The selected specification remains interpretable through its dynamic decomposition into source-specific corrections, trend, seasonal structure, and transfer effects. Its fitted transfer coefficients are quantile-specific rather than uniform: in the representative cutoff, local precipitation and soil moisture have negative coefficients in the lower and median quantile fits and positive coefficients in the 95th quantile fit, while the first GDPC coefficient is comparatively small on the reported scale. The regime diagnostics similarly separate fitted historical behavior into trend, seasonal, and transfer structure, with wider separation between lower and upper fitted quantiles during wetter, high-flow periods than during dry periods.

The framework still depends on the quality and compatibility of the external products and on the availability of forecast covariates at each forecast origin. The five-cutoff evaluation is not a dense continuous hindcast; the limiting step for such an extension is reconstructing version-consistent historical forecast-origin bundles, not only refitting the state-space models. In the present implementation, forecast-window GEFS precipitation and shallow soil-water ensembles enter through deterministic covariate summaries; future work could propagate the full uncertainty in these exogenous forecasts jointly with the river-flow synthesis. Natural extensions include joint treatment of multiple quantiles, richer dynamic structures for the discrepancy process, and spatial generalizations that couple nearby sites or basins.

\section*{Acknowledgments}

This work is part of the Ph.D. dissertation of Antonio De Leon in the statistical science program at the University of California, Santa Cruz. The work was supported in part by the National Science
Foundation grant MMS2050012.

\section*{Code availability}

The reusable estimation routines for the extended dynamic quantile linear model are available through the CRAN R package \texttt{exdqlm}, version 1.1.0 (\url{https://CRAN.R-project.org/package=exdqlm}; package DOI \url{https://doi.org/10.32614/CRAN.package.exdqlm}), and are described in \citet{deleon2026exdqlm}. The clean reproducibility repository for this study, including curated fitting and post-processing workflow code, model-ready staged inputs for the five cutoffs, manuscript-facing outputs, and compact provenance manifests, is available at \url{https://github.com/AntonioAPDL/san-lorenzo-exdqlm-reproducibility}.
The public repository starts from the staged inputs used by the reported analysis and does not bundle raw climate-center retrievals or intermediate covariate-preprocessing workflows. Applying the workflow to a new basin requires staging the corresponding observations, current forecast products, and covariates. Reproducing the retrospective validation design additionally requires a basin-specific, version-consistent archive that aligns historical observations, retrospective products, issued forecast products, forecast-window covariates, product versions, spatial extraction rules, and source-specific forecast horizons.






\clearpage

\appendix

\section{Supplementary Parameter Summaries and Illustrations}
\label{app:extras}

\subsection{Component-Removal Sensitivity Analysis}
\label{app:he3ablation}

To assess how the selected specification depends on its main structural blocks, we refit \texttt{exAL-M-T1} after removing one component at a time. Each reduced model uses the same rolling-origin cutoffs, input data, preprocessing, likelihood, forecast horizons, and selected hyperparameters, so the comparison isolates the specified component removal within this fixed sensitivity design. The labels \texttt{noH1}, \texttt{noH2}, and \texttt{noH3} remove the first, second, and third retained seasonal harmonic pairs, respectively; \texttt{noH3} refers to the retained noninteger frequency \(1/6.8068493\), not to a literal harmonic frequency of 3.

\begin{table*}[htbp]
\centering
\renewcommand{\arraystretch}{1.08}
\begin{threeparttable}
\caption{Component-removal sensitivity analysis for 28-day forecast-window CRPS under the selected \texttt{exAL-M-T1} specification. Lower values are better, and bold marks the best value among the full reference and reduced models within each cutoff.}
\label{tab:he3_ablation_crps}
\begin{tabular*}{\textwidth}{@{\extracolsep{\fill}} >{\ttfamily}l r r r r r}
\toprule
Model & 01/23/2021 & 11/12/2021 & 12/21/2021 & 05/11/2022 & 12/25/2022 \\
\midrule
exAL-M-T1 (full) & \textbf{0.13971} & \textbf{0.04724} & \textbf{0.26045} & \textbf{0.02273} & \textbf{0.53806} \\
exAL-M-T1-noTrend & 1.05159 & 0.72343 & 2.51748 & 0.53257 & 4.05881 \\
exAL-M-noTF & 1.73610 & 1.71409 & 2.32863 & 2.47703 & 2.17280 \\
exAL-M-T1-noH1 & 1.27861 & 1.60142 & 2.83017 & 1.09956 & 4.45887 \\
exAL-M-T1-noH2 & 0.72690 & 1.04699 & 2.96513 & 0.91193 & 4.64709 \\
exAL-M-T1-noH3 & 1.08281 & 1.03562 & 2.58333 & 0.68642 & 4.43879 \\
\addlinespace[2pt]
\multicolumn{6}{l}{\textit{Raw forecast products (reference only)}} \\
RAW-GLOFAS & 0.40366 & 0.16957 & 0.68246 & 0.27227 & 1.56006 \\
\bottomrule
\end{tabular*}
\begin{tablenotes}
\item \textit{Note:} The \texttt{noH3} row removes the retained noninteger seasonal harmonic with frequency \(1/6.8068493\).
\end{tablenotes}
\end{threeparttable}
\end{table*}

\begin{table*}[htbp]
\centering
\renewcommand{\arraystretch}{1.08}
\begin{threeparttable}
\caption{Component-removal sensitivity analysis over the common eight-day NWS forecast horizon. Lower values are better, and bold marks the best value among the full reference and reduced models within each cutoff.}
\label{tab:he3_ablation_crps_nws_horizon}
\begin{tabular*}{\textwidth}{@{\extracolsep{\fill}} >{\ttfamily}l r r r r r}
\toprule
Model & 01/23/2021 & 11/12/2021 & 12/21/2021 & 05/11/2022 & 12/25/2022 \\
\midrule
exAL-M-T1 (full) & \textbf{0.22691} & \textbf{0.05858} & \textbf{0.59445} & \textbf{0.02567} & \textbf{0.59070} \\
exAL-M-T1-noTrend & 1.49490 & 0.77464 & 3.38697 & 0.64627 & 3.40333 \\
exAL-M-noTF & 1.79777 & 1.69066 & 2.18697 & 2.45857 & 2.30863 \\
exAL-M-T1-noH1 & 1.66963 & 1.69424 & 3.70227 & 1.14270 & 3.79228 \\
exAL-M-T1-noH2 & 1.11122 & 1.14286 & 3.80753 & 0.99851 & 3.92807 \\
exAL-M-T1-noH3 & 1.53419 & 1.09181 & 3.45450 & 0.78219 & 3.75262 \\
\addlinespace[2pt]
\multicolumn{6}{l}{\textit{Raw forecast products (reference only)}} \\
RAW-GLOFAS & 0.57251 & 0.20670 & 1.29127 & 0.32274 & 1.48188 \\
RAW-NWS & 0.83037 & 1.37192 & 0.28118 & 0.28366 & 0.55680 \\
\bottomrule
\end{tabular*}
\begin{tablenotes}
\item \textit{Note:} This table restricts every row to forecast leads 1--8, the common daily horizon available for the NWS comparison.
\end{tablenotes}
\end{threeparttable}
\end{table*}

The full reference model has lower CRPS than each reduced model at all five cutoffs under both the 28-day and eight-day horizon contracts. This component-removal sensitivity analysis supports retaining the transfer, trend, and multi-frequency seasonal blocks in the selected specification.

\subsection{Supplementary Source-Specific Shape and Scale Parameters}
\label{app:sourceparams}

\begin{table*}[htbp]
\centering
\begin{threeparttable}
\caption{Posterior Medians and 95\% Credible Intervals for the Source-Specific Skewness Parameters $\gamma_j(\tau)$ at the Representative December 25, 2022 Cutoff}
\label{tab:gamma_sigma_intervals1}
\begin{tabular*}{\textwidth}{@{\extracolsep{\fill}} l
 S[table-format=-1.5] l
 S[table-format=-1.5] l
 S[table-format=-1.5] l
@{}}
\toprule
 & \multicolumn{2}{c}{\textbf{USGS}} & \multicolumn{2}{c}{\textbf{GLOFAS}} & \multicolumn{2}{c}{\textbf{NWS}} \\
\cmidrule(lr){2-3}\cmidrule(lr){4-5}\cmidrule(lr){6-7}
\textbf{Quantile} & \textbf{Median} & \textbf{95\% CI} & \textbf{Median} & \textbf{95\% CI} & \textbf{Median} & \textbf{95\% CI} \\
\midrule
05th & 0.84518 & $[0.83064,\ 0.85881]$ & 0.97496 & $[0.96029,\ 0.98916]$ & 0.48124 & $[0.47043,\ 0.49255]$ \\
20th & 0.09424 & $[0.08194,\ 0.10702]$ & 0.33544 & $[0.32203,\ 0.34881]$ & -0.00668 & $[-0.00924,\ -0.00413]$ \\
35th & -0.02763 & $[-0.03298,\ -0.02234]$ & 0.16221 & $[0.15243,\ 0.17236]$ & -0.04280 & $[-0.04707,\ -0.03841]$ \\
50th & -0.13196 & $[-0.14005,\ -0.12322]$ & 0.06336 & $[0.05671,\ 0.07050]$ & -0.13248 & $[-0.13938,\ -0.12523]$ \\
65th & -0.31071 & $[-0.32244,\ -0.29846]$ & 0.01977 & $[0.01512,\ 0.02466]$ & -0.27078 & $[-0.28067,\ -0.26029]$ \\
80th & -0.68543 & $[-0.70270,\ -0.66835]$ & -0.05100 & $[-0.06204,\ -0.04053]$ & -0.49724 & $[-0.51116,\ -0.48282]$ \\
95th & -2.13801 & $[-2.16788,\ -2.10810]$ & -0.72784 & $[-0.73970,\ -0.71634]$ & -1.49200 & $[-1.51264,\ -1.47052]$ \\
\bottomrule
\end{tabular*}
\begin{tablenotes}
\item \textit{Note:} Posterior medians and 95\% credible intervals $[{\rm Q2.5}, {\rm Q97.5}]$ for the source-specific skewness parameters $\gamma_j(\tau)$ in the exAL observation model at the representative December 25, 2022 cutoff. The quantile labels 05th through 95th correspond to the seven fitted quantile models. These summaries are included as supplementary appendix support rather than as primary forecast-validation evidence.
\end{tablenotes}
\end{threeparttable}
\end{table*}

\begin{table*}[htbp]
\centering
\begin{threeparttable}
\caption{Posterior Medians and 95\% Credible Intervals for the Source-Specific Scale Parameters $\sigma_j(\tau)$ at the Representative December 25, 2022 Cutoff}
\label{tab:gamma_sigma_intervals2}
\begin{tabular*}{\textwidth}{@{\extracolsep{\fill}} l
 S[table-format=1.5] l
 S[table-format=1.5] l
 S[table-format=1.5] l
@{}}
\toprule
 & \multicolumn{2}{c}{\textbf{USGS}} & \multicolumn{2}{c}{\textbf{GLOFAS}} & \multicolumn{2}{c}{\textbf{NWS}} \\
\cmidrule(lr){2-3}\cmidrule(lr){4-5}\cmidrule(lr){6-7}
\textbf{Quantile} & \textbf{Median} & \textbf{95\% CI} & \textbf{Median} & \textbf{95\% CI} & \textbf{Median} & \textbf{95\% CI} \\
\midrule
05th & 0.02923 & $[0.02884,\ 0.02965]$ & 0.02078 & $[0.02052,\ 0.02104]$ & 0.02829 & $[0.02790,\ 0.02866]$ \\
20th & 0.05872 & $[0.05793,\ 0.05954]$ & 0.03951 & $[0.03900,\ 0.04001]$ & 0.06574 & $[0.06476,\ 0.06663]$ \\
35th & 0.07519 & $[0.07411,\ 0.07621]$ & 0.04833 & $[0.04767,\ 0.04895]$ & 0.09353 & $[0.09219,\ 0.09500]$ \\
50th & 0.08073 & $[0.07960,\ 0.08190]$ & 0.05251 & $[0.05185,\ 0.05324]$ & 0.10690 & $[0.10534,\ 0.10837]$ \\
65th & 0.07686 & $[0.07580,\ 0.07800]$ & 0.05317 & $[0.05247,\ 0.05392]$ & 0.10780 & $[0.10630,\ 0.10926]$ \\
80th & 0.06768 & $[0.06681,\ 0.06864]$ & 0.04415 & $[0.04353,\ 0.04475]$ & 0.09458 & $[0.09338,\ 0.09584]$ \\
95th & 0.04631 & $[0.04581,\ 0.04681]$ & 0.02315 & $[0.02286,\ 0.02345]$ & 0.05602 & $[0.05533,\ 0.05674]$ \\
\bottomrule
\end{tabular*}
\begin{tablenotes}
\item \textit{Note:} Posterior medians and 95\% credible intervals $[{\rm Q2.5}, {\rm Q97.5}]$ for the source-specific scale parameters $\sigma_j(\tau)$ at the representative December 25, 2022 cutoff. These summaries are included as supplementary appendix support rather than as primary forecast-validation evidence.
\end{tablenotes}
\end{threeparttable}
\end{table*}

Tables~\ref{tab:gamma_sigma_intervals1} and \ref{tab:gamma_sigma_intervals2} provide summaries of the source-specific skewness (\(\gamma\)) and scale (\(\sigma\)) parameters in the exAL observation model, reported for the December 25, 2022 cutoff. These are likelihood-parameter summaries, not synthesis weights.

\subsection{Univariate Transfer-Active Predictive Synthesis}
\label{app:historicalsynth}

\begin{figure}[htbp]
\centering
\includegraphics[width=450pt]{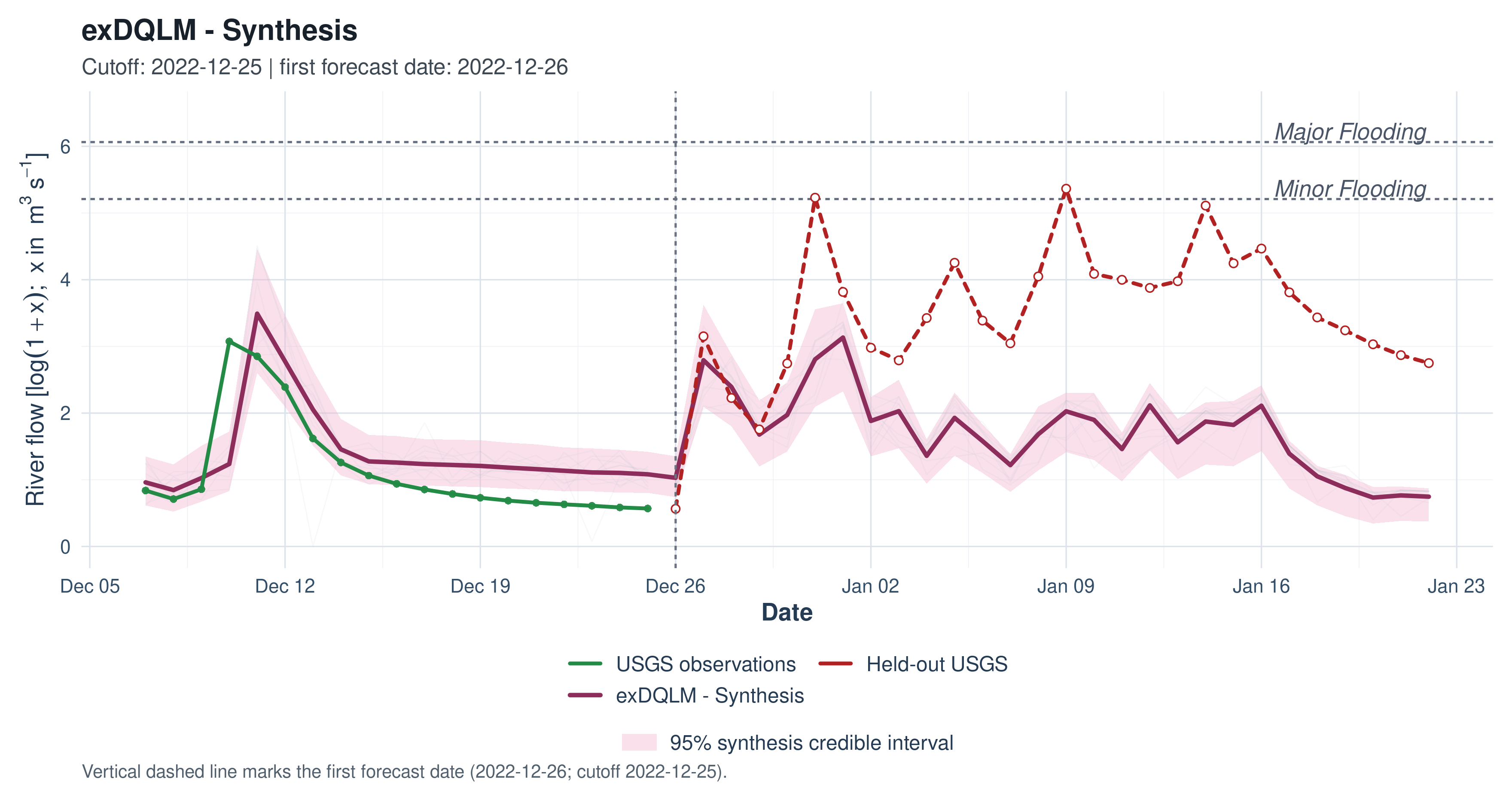}
\caption{
Univariate transfer-active predictive distribution for river flow on the $\log(1+x)$ scale, with $x$ measured in m$^3$/s, at the December 25, 2022 forecast origin. Observed measurements before the origin are shown in green, and held-out USGS verification after the origin is shown in red. The pink band is the central 95\% posterior predictive credible band from the reference synthesized distribution, and the dark center line summarizes the corresponding synthesized center. The reference uses the USGS observation channel and transfer covariates, but excludes retrospective-product and forecast-product source channels.
}
\label{fig:synth2}
\end{figure}

Figure~\ref{fig:synth2} provides a univariate transfer-active reference for the December 25, 2022 cutoff. It is included to contrast the multivariate operational synthesis in Figure~\ref{fig:synth1} with a simpler reference that excludes retrospective-product and forecast-product source channels while retaining the transfer/covariate component.

\subsection{Additional Cutoff-Specific Predictive Synthesis Panels}
\label{app:cutoffsynthpanels}

The main text shows the December 25, 2022 synthesis as a representative selected-model illustration. For completeness, this appendix reports the corresponding multivariate synthesis panels for the other rolling-origin cutoffs. Each panel uses the same exAL-M-T1 synthesis specification as Figure~\ref{fig:synth1} and includes the retrospective and forecast-product references used at that cutoff.

\begin{figure}[p]
\centering
\includegraphics[width=450pt]{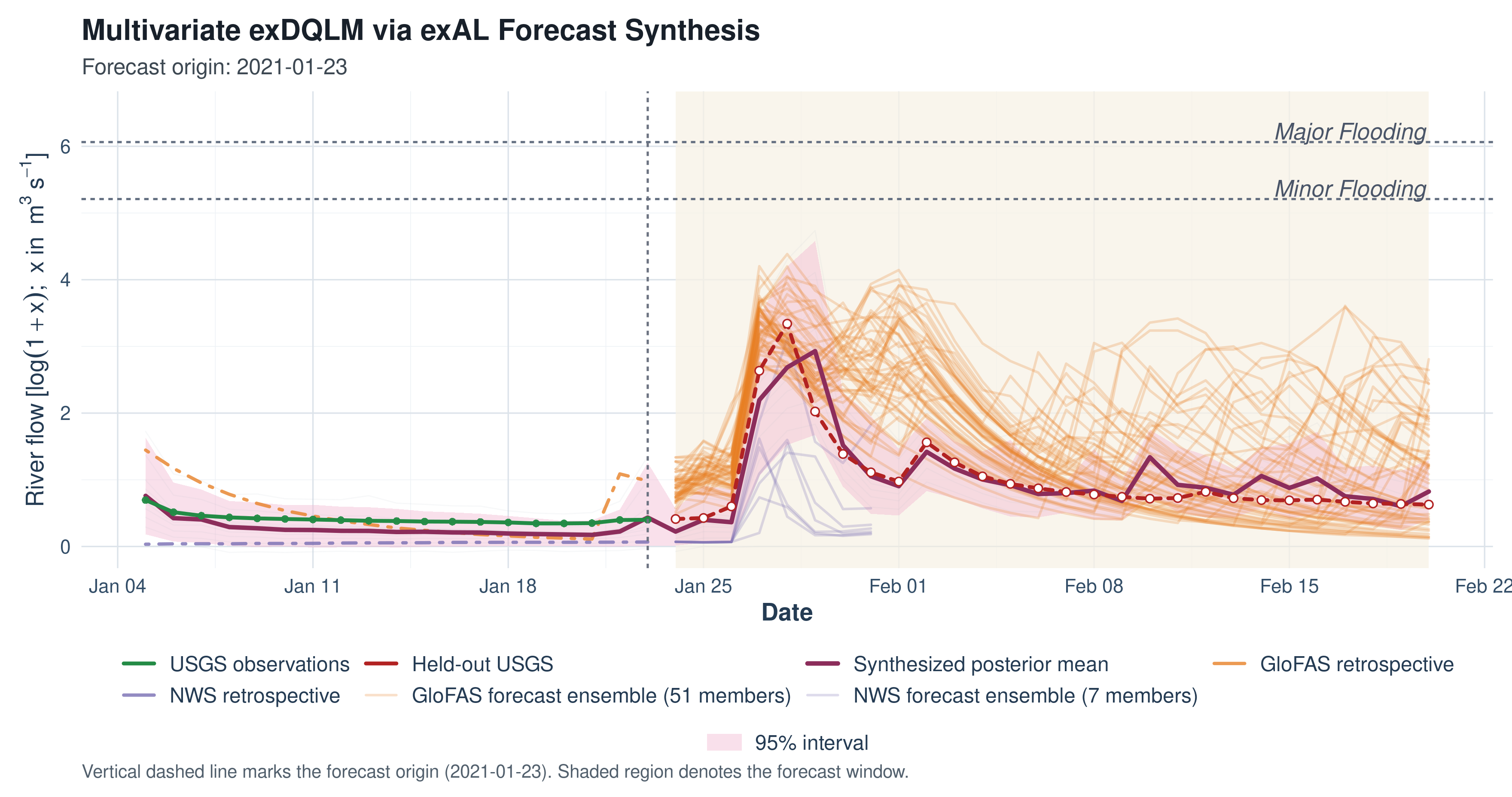}
\caption{
Predictive synthesis panel for the January 23, 2021 cutoff under the selected exAL-M-T1 specification. Observed USGS measurements before the origin are shown in green, and held-out USGS verification after the origin is shown in red. The pink band is the central 95\% posterior predictive credible band from the synthesized distribution, the dark center line summarizes the corresponding synthesized center, and the thin overlaid lines show the displayed synthesized quantiles. Raw retrospective and forecast products are included as reference overlays for the cutoff.
}
\label{fig:synthsupport_20210123}
\end{figure}

\begin{figure}[p]
\centering
\includegraphics[width=450pt]{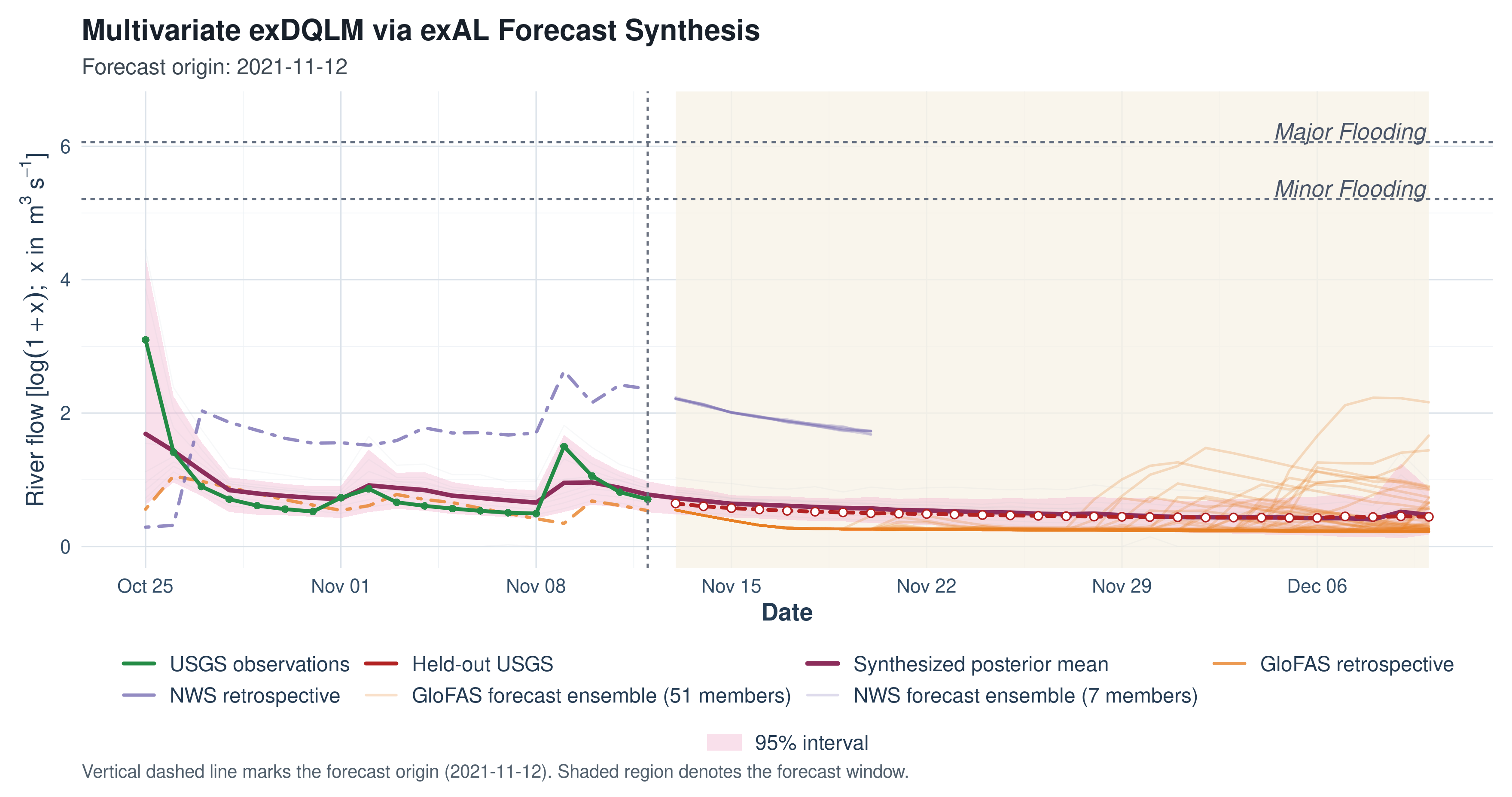}
\caption{
Predictive synthesis panel for the November 12, 2021 cutoff under the selected exAL-M-T1 specification. Observed USGS measurements before the origin are shown in green, and held-out USGS verification after the origin is shown in red. The pink band is the central 95\% posterior predictive credible band from the synthesized distribution, the dark center line summarizes the corresponding synthesized center, and the thin overlaid lines show the displayed synthesized quantiles. Raw retrospective and forecast products are included as reference overlays for the cutoff.
}
\label{fig:synthsupport_20211112}
\end{figure}

\begin{figure}[p]
\centering
\includegraphics[width=450pt]{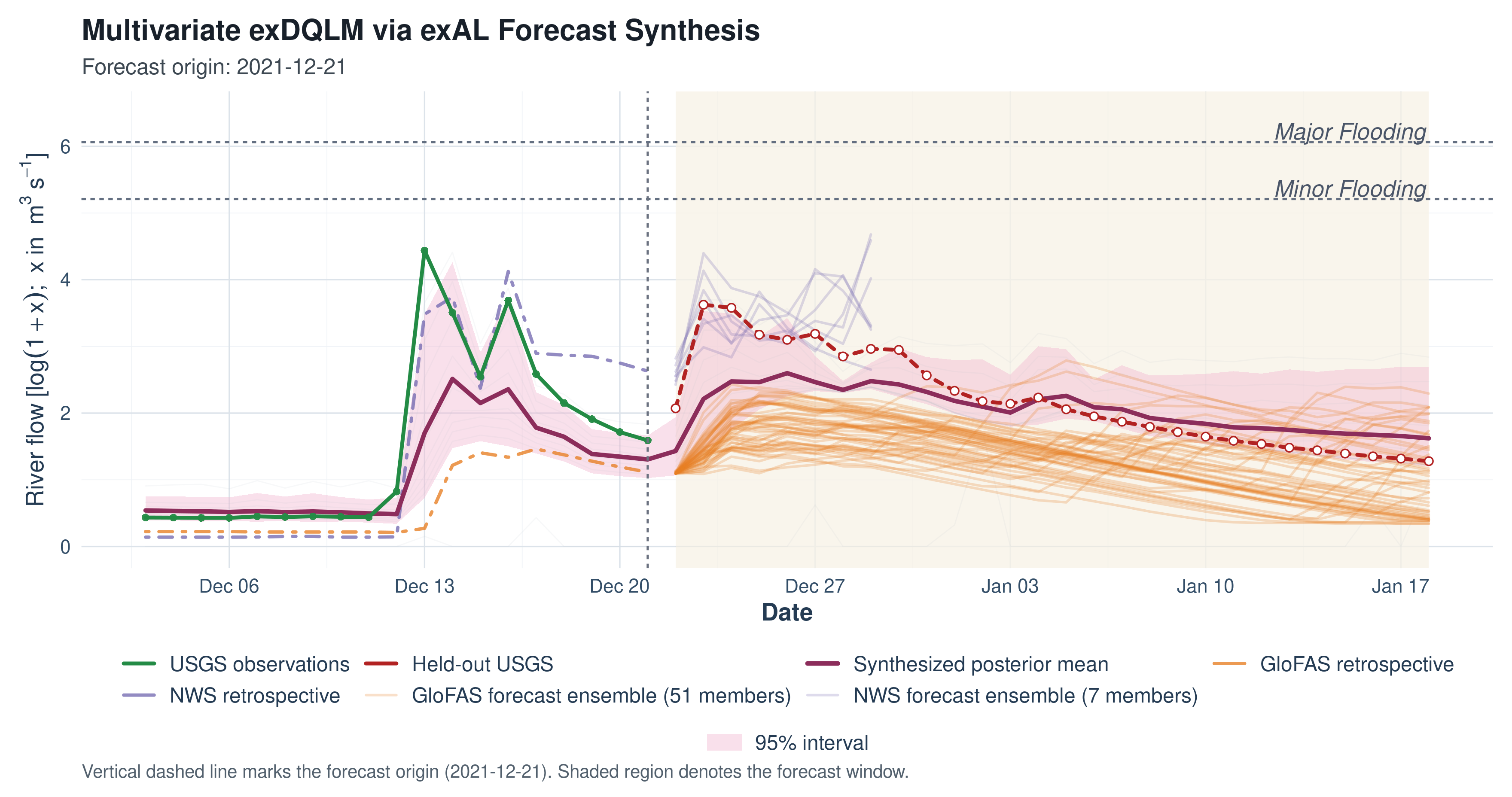}
\caption{
Predictive synthesis panel for the December 21, 2021 cutoff under the selected exAL-M-T1 specification. Observed USGS measurements before the origin are shown in green, and held-out USGS verification after the origin is shown in red. The pink band is the central 95\% posterior predictive credible band from the synthesized distribution, the dark center line summarizes the corresponding synthesized center, and the thin overlaid lines show the displayed synthesized quantiles. Raw retrospective and forecast products are included as reference overlays for the cutoff.
}
\label{fig:synthsupport_20211221}
\end{figure}

\begin{figure}[p]
\centering
\includegraphics[width=450pt]{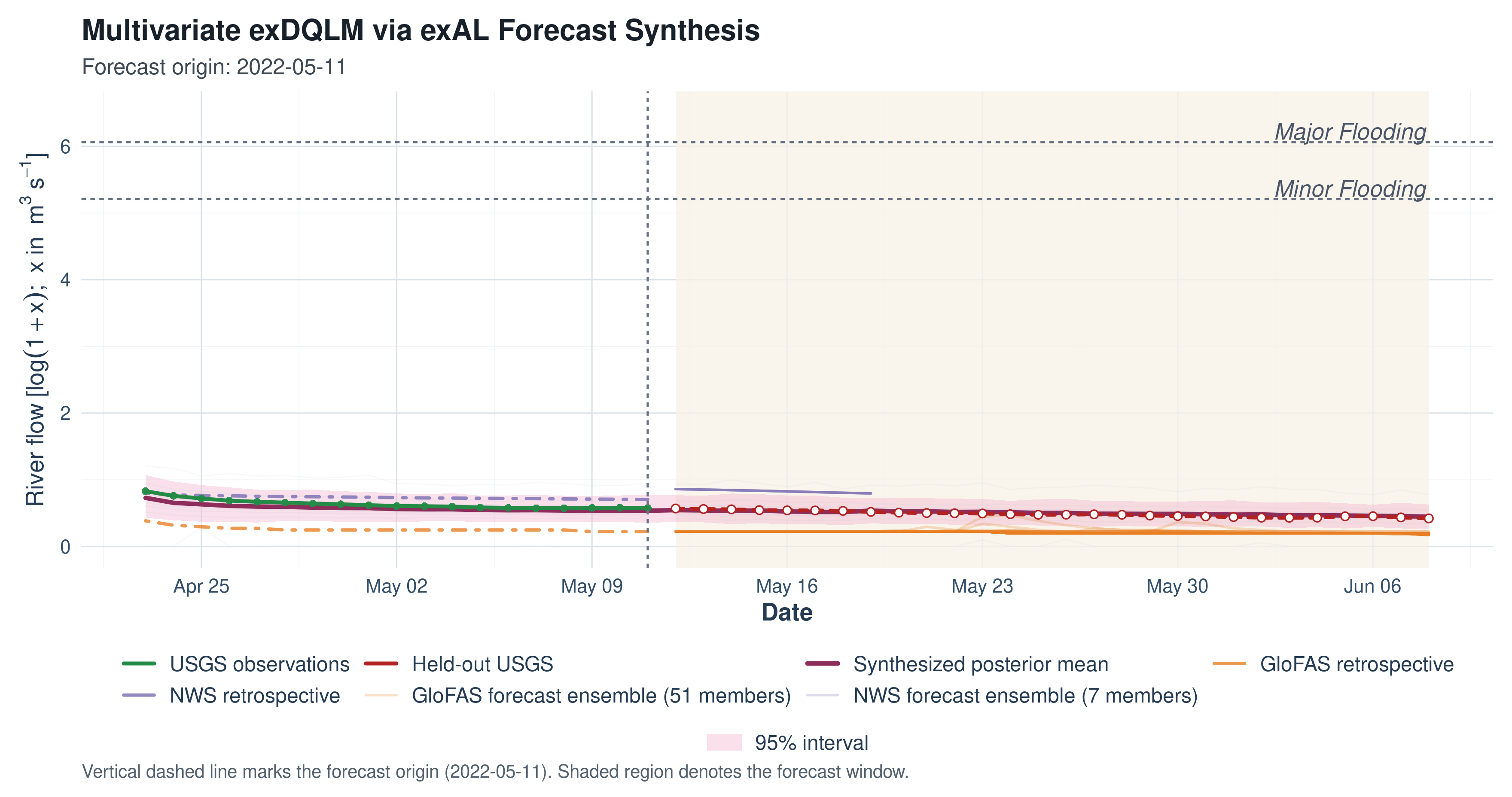}
\caption{
Predictive synthesis panel for the May 11, 2022 cutoff under the selected exAL-M-T1 specification. Observed USGS measurements before the origin are shown in green, and held-out USGS verification after the origin is shown in red. The pink band is the central 95\% posterior predictive credible band from the synthesized distribution, the dark center line summarizes the corresponding synthesized center, and the thin overlaid lines show the displayed synthesized quantiles. Raw retrospective and forecast products are included as reference overlays for the cutoff.
}
\label{fig:synthsupport_20220511}
\end{figure}

\section{Markov Chain Monte Carlo Algorithms}

The appendix algorithms use the same state-space notation as Subsection~\ref{subsec:fullmodel}. To keep the displays compact, they work with source-indexed observation vectors during the pre-cutoff observation window and with source-indexed forecast-product vectors during the forecast window. The role of the selector vector is therefore the same throughout: before the cutoff it is \(\mathbf{h}_{t,j}\), whereas after the cutoff it becomes the forecast-window selector \(\mathbf{e}_{T+k,j}\) for the active source \(j\). When the main text distinguishes individual forecast members through \(y_T^{j,i}(k)\), the appendix uses the corresponding source-indexed forecast quantity \(y_{T+k}^{j}\) because the algorithms only need the active source channel at each lead time.

\begin{algorithm}
\caption{MCMC Algorithm}\label{alg1}
\begin{algorithmic}

\State \textbf{Notation:}
\State  \(
A^j = A(\gamma^j;p_0), \quad B^j = B(\gamma^j;p_0), \quad C^j = C(\gamma^j;p_0), \text{ as defined in Subsection~\ref{subsec:exal}}
\)
\State Let \(\mathcal{J}_0=\{0\}\cup\mathcal{J}\), where \(j=0\) denotes the observation channel and \(j\in\mathcal{J}\) denotes an external source. During the pre-cutoff observation window, write \(y_t^0 \equiv y_t^o\) and \(y_t^j \equiv z_t^j\) for \(j\in\mathcal{J}\), with selector vector \(\mathbf{h}_{t,j}\). During the forecast window, write \(y_{T+k}^{j}\) for the active source-indexed forecast quantity from source \(j\) at lead \(k\), with selector \(\mathbf{e}_{T+k,j}\).
\State \textbf{Step 0: Set initial values} and sample size $N$.
\State Initialize $\{(v^j)^{(0)}_{1:T}, (s^j)^{(0)}_{1:T}, \boldsymbol{\eta}^{(0)}_{1:T}, (\sigma^j)^{(0)}, (\gamma^j)^{(0)}\}_{j\in\mathcal{J}_0}$
\For{$n = 1$ to $N$}
    \State \textbf{Step 1: Update $\boldsymbol{\eta}_t$}
    \State Use Algorithm~\ref{alg:Alg2}
    \For{$j \in \mathcal{J}_0$}
        \State \textbf{Step 2: Update $v^j_t$}
        \For{$t = 1$ to $T$}
            \State Sample $(v^j_t)^{(n)} \sim \mathcal{GIG}(a^j_t, b^j_t, c^j_t)$
            \State $a^j_t = \frac{1}{2}$
            \State $b^j_t = \frac{1}{(\sigma^j)^{(n-1)}}\left[ \frac{((A^j)^2)^{(n-1)}}{(B^j)^{(n-1)}}+2\right]$
            \State $c^j_t = \frac{1}{(\sigma^j)^{(n-1)} (B^j)^{(n-1)}}\left[y_t^j -\mathbf{h}_{t,j}' \boldsymbol{\eta}^{(n)}_t -(C^j)^{(n-1)} (\sigma^j)^{(n-1)} |(\gamma^j)^{(n-1)}| (s^j_t)^{(n-1)} \right]^2$
        \EndFor
        \If{$j \in \mathcal{J}$}
        \For{$k = 1$ to $K_{(j)}$}
            \State Sample $(v^j_{T+k})^{(n)} \sim \mathcal{GIG}(a^j_{T+k}, b^j_{T+k}, c^j_{T+k})$
            \State $a^j_{T+k} = \frac{1}{2}$
            \State $b^j_{T+k} = \frac{1}{(\sigma^j)^{(n-1)}}\left[ \frac{((A^j)^2)^{(n-1)}}{(B^j)^{(n-1)}}+2\right]$
            \State $c^j_{T+k} = \frac{1}{(\sigma^j)^{(n-1)} (B^j)^{(n-1)}}\left[y_{T+k}^j -\mathbf{e}_{T+k,j}' \boldsymbol{\eta}^{(n)}_{T+k} -(C^j)^{(n-1)} (\sigma^j)^{(n-1)} |(\gamma^j)^{(n-1)}| (s^j_{T+k})^{(n-1)} \right]^2$
        \EndFor
        \EndIf

        \State \textbf{Step 3: Update $s^j_t$}
        \For{$t = 1$ to $T$}
            \State Sample $(s^j)^{(n)}_t \sim \mathcal{N}^+(m^j_t, \lambda^j_t)$
            \State $\lambda^j_t =  \left[\frac{\big((C^j)^{(n-1)}(\gamma^j)^{(n-1)} \big)^2(\sigma^j)^{(n-1)}}{(B^j)^{(n-1)}(v^j_t)^{(n)}}+1\right]$
            \State $m^j_t =  \lambda^j_t \left[\frac{(C^j)^{(n-1)}|(\gamma^j)^{(n-1)}|(y_t^j -\mathbf{h}_{t,j}' \boldsymbol{\eta}^{(n)}_t - (A^j)^{(n-1)} (v^j_t)^{(n)})}{(B^j)^{(n-1)}(v^j_t)^{(n)}}\right]$
        \EndFor

        \State \textbf{Step 4: Update $\sigma^j$}
        \State Sample $(\sigma^j)^{(n)} \sim \mathcal{GIG}(a^j_t, b^j_t, c^j_t)$
        \State $a^j_t = -(1.5T + a_{\sigma^j})$
        \State $b^j_t = \sum_{t=1}^T \frac{1}{(B^j)^{(n-1)} (v^j_t)^{(n)}} \left((C^j)^{(n-1)} |(\gamma^j)^{(n-1)}| (s^j_t)^{(n)} \right)^2$
        \State $c^j_t = \sum_{t=1}^T \frac{1}{(B^j)^{(n-1)} (v^j_t)^{(n)}}\big(y_t^j -\mathbf{h}_{t,j}' \boldsymbol{\eta}^{(n)}_t - (A^j)^{(n-1)} (v^j_t)^{(n)}\big)^2 + 2 \sum_{t=1}^T  (v^j_t)^{(n)} + 2 b_{\sigma^j}$

        \State \textbf{Step 5: Update $\gamma^j$}
        \State Implement Metropolis-Hastings (MH)
    \EndFor
\EndFor

\end{algorithmic}
\end{algorithm}

\begin{algorithm}
\caption{MCMC - Forward Filtering Backward Sampling (FFBS)}
\label{alg:Alg2}
\begin{algorithmic}

\State \textbf{Notation:}
\State  During the pre-cutoff observation window, \(\mathbf{H}_t = [\mathbf{h}_{t,0}, \mathbf{h}_{t,1}, \ldots, \mathbf{h}_{t,J}]\), \(\mathbf{Y}_t = [y_t^o, z_t^1, \ldots, z_t^J]'\), and \(\mathbf{D}_t, \boldsymbol{\Lambda}_t\) are as defined in Subsection~\ref{subsec:fullmodel}.
\State  Equivalently, \(y_t^0 \equiv y_t^o\) and \(y_t^j \equiv z_t^j\) for \(j\in\mathcal{J}\), so the filtering recursion can be written in source-indexed form.
\State $\mathbf{M}_t= \mathbf{C} \odot  \mathbf{\sigma} \odot | \mathbf{\gamma} | \odot \mathbf{s}_t + \mathbf{A} \odot \mathbf{v}_t$, \quad $\mathbf{O}_t= \sqrt{\mathbf{B} \odot \mathbf{v}_t} \odot Diag
(\mathbf{\sigma}) \sqrt{\mathbf{B} \odot \mathbf{v}_t} \odot \mathbf{\sigma}$
\State  \(
\mathbf{A} = \big[ A(\gamma^j;p_0)   \big], \quad \mathbf{B} = \big[  B(\gamma^j;p_0)  \big], \quad \mathbf{C} = \big[ C(\gamma^j;p_0)   \big] \), \text{ as defined in Subsection~\ref{subsec:exal}}
\State \( \mathbf{v}_t = \big[  v^j_t   \big], \quad  \mathbf{s}_t = \big[
s^j_t  \big] , \quad \mathbf{\sigma} = \big[ \sigma^j  \big]
\)
\State \(\odot\) denotes the Hadamard product.
\State \textbf{Step 1: Forward Filtering}
\For{$t = 1$ to $T$}
    \State Compute Prior moments:
    \State $\mathbf{a}_t = \mathbf{D}_t \mathbf{m}_{t-1}$
    \State $\mathbf{R}_t = \mathbf{D}_t \mathbf{C}_{t-1} \mathbf{D}_t' +  \boldsymbol{\Lambda}_t$

    \State One-step-ahead forecast:
    \State $\mathbf{f}_t = \mathbf{H}_t' \mathbf{a}_t + \mathbf{M}_t$
    \State $\mathbf{Q}_t = \mathbf{H}_t' \mathbf{R}_t \mathbf{H}_t + \mathbf{O}_t$

    \State Compute Posterior moments:
    \State $\mathbf{m}_t = \mathbf{a}_t + \mathbf{R}_t \mathbf{H}_t \mathbf{Q}_t^{-1}(\mathbf{Y}_t - \mathbf{f}_t)$
    \State $\mathbf{C}_t = \mathbf{R}_t - \mathbf{R}_t \mathbf{H}_t \mathbf{Q}_t^{-1} \mathbf{H}_t' \mathbf{R}_t$
\EndFor

\State \textbf{Step 2: Backward Sampling}
\For{$t = T$ down to $0$}
    \If{$t = T$}
        \State Sample $\boldsymbol{\eta}_T | D_T \sim \mathcal{N}(\mathbf{m}_T, \mathbf{C}_T)$
    \Else
        \State Compute Retrospective moments:
        \State $\mathbf{m}_t^s = \mathbf{m}_t + \mathbf{C}_t \mathbf{D}_{t+1}' \mathbf{R}_{t+1}^{-1} (\boldsymbol{\eta}_{t+1} - \mathbf{a}_{t+1})$
        \State $\mathbf{C}_t^s = \mathbf{C}_t - \mathbf{C}_t \mathbf{D}_{t+1}' \mathbf{R}_{t+1}^{-1} \mathbf{D}_{t+1} \mathbf{C}_t$

        \State Sample $\boldsymbol{\eta}_t | D_T \sim \mathcal{N}(\mathbf{m}_t^s, \mathbf{C}_t^s)$
    \EndIf
\EndFor

\end{algorithmic}
\end{algorithm}

 \clearpage

\section{Variational Bayes Algorithms}

\begin{algorithm}
\caption{Variational Bayes (VB) Algorithm}
\label{alg:Alg3}
\begin{algorithmic}

\State \textbf{Notation:}
\State  $\left\langle g(\xi) \right\rangle = \mathbb{E}_\xi\left[ g(\xi) \right]$
\State  \(
A^j = A(\gamma^j;p_0), \quad B^j = B(\gamma^j;p_0), \quad C^j = C(\gamma^j;p_0), \text{ as defined in Subsection~\ref{subsec:exal}}
\)
\State  Let \(\mathcal{J}_0=\{0\}\cup\mathcal{J}\). During the pre-cutoff observation window, let \(y_t^0 \equiv y_t^o\) and \(y_t^j \equiv z_t^j\) for \(j\in\mathcal{J}\), with selector \(\mathbf{h}_{t,j}\). This is the same source-indexed notation used in Algorithm~\ref{alg1}.
\State \textbf{Step 0: Set initial values}
\State Choose \(k_{\max}\), minimum update counts, and convergence tolerances; set \(k=0\)

\While{\(k < k_{\max}\) and convergence criteria are not met}
    \State \(k \leftarrow k+1\)
    \State \textbf{Step 1: Update $r(\boldsymbol{\eta}_t)=\mathcal{N}(m^{(k)}_t,C^{(k)}_t)$}
    \State Use Algorithm~\ref{alg:Alg4}

    \For{$j \in \mathcal{J}_0$}
        \State \textbf{Step 2: Update $r(v^j_t)=\mathcal{GIG}(a_t^{j,(k)},b_t^{j,(k)},c_t^{j,(k)})$}
        \For{$t = 1$ to $T$}
            \State $a_t^{j,(k)} = \frac{1}{2}$
            \State $b_t^{j,(k)} = \left\langle \frac{A^2}{\sigma^j B^j} \right\rangle^{(k-1)} + 2 \left\langle \frac{1}{\sigma^j} \right\rangle^{(k-1)}$
            \State $c_t^{j,(k)} = \left\langle \frac{1}{\sigma^j B^j} \right\rangle^{(k-1)} \Big[ (y_t^j)^2 - 2y_t^j \left\langle  \mathbf{h}_{t,j}' \boldsymbol{\eta}_t \right\rangle^{(k)} + \left\langle (  \mathbf{h}_{t,j}' \boldsymbol{\eta}_t )^2 \right\rangle^{(k)} \Big]$
            \State $\quad \quad -2 \left\langle s^j_t \right\rangle^{(k-1)} \left\langle \frac{C^j|\gamma^j|}{B^j} \right\rangle^{(k-1)} \Big[ y_t^j - \left\langle  \mathbf{h}_{t,j}' \boldsymbol{\eta}_t \right\rangle^{(k)} \Big]$
            \State $\quad \quad + \left\langle (s^j_t)^2 \right\rangle^{(k-1)} \left\langle \frac{(C^j)^2\sigma^j|\gamma^j|^2}{B^j} \right\rangle^{(k-1)}$
        \EndFor

        \State \textbf{Step 3: Update $r(s^j_t)=\mathcal{N}^+(\eta^{j,(k)},\lambda^{j,(k)})$}
        \For{$t = 1$ to $T$}
            \State $\lambda^{j,(k)} = \Big[ 1+\left\langle \frac{(C^j)^2 \sigma^j |(\gamma^j)|^2}{B^j} \right\rangle^{(k-1)} \left\langle \frac{1}{v^j_t} \right\rangle^{(k)} \Big]^{-1}$
            \State $\eta^{j,(k)} = \lambda^{j,(k)}  \Big[ \left\langle \frac{C^j |\gamma^j|}{B^j} \right\rangle^{(k-1)} \left\langle \frac{1}{v^j_t} \right\rangle^{(k)} \Big( y_t^j - \left\langle  \mathbf{h}_{t,j}' \boldsymbol{\eta}_t \right\rangle^{(k)} \Big) - \left\langle \frac{C^j |\gamma^j| A^j}{B^j} \right\rangle^{(k-1)} \Big]$
        \EndFor

        \State \textbf{Step 4: Update $r(\gamma^j, \sigma^j )$}
        \State Approximate via VB-LD, as described in Subsection~\ref{subsec:postinf}
    \EndFor
    \State \textbf{Step 5: Check convergence}
    \State Compute \(\mathcal{L}^{(k)}\), \(\Delta_{\mathcal{L}}^{(k)}=|\mathcal{L}^{(k)}-\mathcal{L}^{(k-1)}|\), and the relative ELBO change
    \State Declare convergence after required update iterations when ELBO,
    \Statex \hspace{\algorithmicindent} state, scale, and skewness changes are below their tolerances
\EndWhile

\end{algorithmic}
\end{algorithm}

\begin{algorithm}
\caption{VB-Filtering and VB-Smoothing}
\label{alg:Alg4}
\begin{algorithmic}

\State \textbf{Notation:}
\State $\left\langle g(\xi) \right\rangle = \mathbb{E}_\xi\left[ g(\xi) \right]$
\State  During the pre-cutoff observation window, \(\mathbf{H}_t = [\mathbf{h}_{t,0}, \mathbf{h}_{t,1}, \ldots, \mathbf{h}_{t,J}]\), \(\mathbf{Y}_t = [y_t^o, z_t^1, \ldots, z_t^J]'\), and \(\mathbf{D}_t, \boldsymbol{\Lambda}_t\) are as defined in Subsection~\ref{subsec:fullmodel}.
\State  Equivalently, let \(y_t^0 \equiv y_t^o\) and \(y_t^j \equiv z_t^j\) for \(j\in\mathcal{J}\). This algorithm is the variational analogue of Algorithm~\ref{alg:Alg2}.
\State $\mathbf{M}_t= \mathbf{C} \odot  \mathbf{\sigma} \odot | \mathbf{\gamma} | \odot \mathbf{s}_t + \mathbf{A} \odot \mathbf{v}_t$, \quad $\mathbf{O}_t= \sqrt{\mathbf{B} \odot \mathbf{v}_t} \odot Diag
(\mathbf{\sigma}) \sqrt{\mathbf{B} \odot \mathbf{v}_t} \odot \mathbf{\sigma}$
\State  \(
\mathbf{A} = \big[ \left\langle A(\gamma^j;p_0)  \right\rangle \big], \quad \mathbf{B} = \big[ \left\langle B(\gamma^j;p_0)  \right\rangle \big], \quad \mathbf{C} = \big[ \left\langle C(\gamma^j;p_0)  \right\rangle \big] \), \text{ as defined in Subsection~\ref{subsec:exal}}
\State \( \mathbf{v}_t = \big[ \left\langle v^j_t  \right\rangle \big], \quad  \mathbf{s}_t = \big[ \left\langle
s^j_t  \right\rangle \big] , \quad \mathbf{\sigma} = \big[ \left\langle \sigma^j  \right\rangle \big]
\)
\State \(\odot\) denotes the Hadamard product.
\State \textbf{Step 1: Forward Filtering}
\For{$t = 1$ to $T$}
    \State Compute Prior moments:
    \State $\mathbf{a}_t = \mathbf{D}_t \mathbf{m}_{t-1}$
    \State $\mathbf{R}_t = \mathbf{D}_t \mathbf{C}_{t-1} \mathbf{D}_t' +  \boldsymbol{\Lambda}_t$

    \State One-step-ahead forecast:
    \State $\mathbf{f}_t = \mathbf{H}_t'\mathbf{a}_t + \mathbf{M}_t$
    \State $\mathbf{Q}_t = \mathbf{H}_t'\mathbf{R}_t\mathbf{H}_t + \mathbf{O}_t$

    \State Compute Posterior moments:
    \State $\mathbf{m}_t = \mathbf{a}_t + \mathbf{R}_t \mathbf{H}_t \mathbf{Q}_t^{-1}(\mathbf{Y}_t - \mathbf{f}_t)$
    \State $\mathbf{C}_t = \mathbf{R}_t - \mathbf{R}_t \mathbf{H}_t \mathbf{Q}_t^{-1} \mathbf{H}_t' \mathbf{R}_t$
\EndFor

\State \textbf{Step 2: Backward Smoothing}
\For{$t = T$ down to $1$}
    \If{$t = T$}
        \State Set $r^{(k+1)}(\boldsymbol{\eta}_T) := \mathcal{N}(\mathbf{m}_T^s, \mathbf{C}_T^s)$
        \State $\mathbf{m}_T^s = \mathbf{m}_T$
        \State $\mathbf{C}_T^s = \mathbf{C}_T$
    \Else
        \State Compute Smoothing moments:
        \State $r^{(k+1)}(\boldsymbol{\eta}_t) := \mathcal{N}(\mathbf{m}_t^s, \mathbf{C}_t^s)$
        \State $B_t = \mathbf{C}_t \mathbf{D}_{t+1}' \mathbf{R}_{t+1}^{-1}$
        \State $\mathbf{m}_t^s = \mathbf{m}_t + \mathbf{B}_t (\mathbf{m}_{t+1}^s - \mathbf{a}_{t+1})$
        \State $\mathbf{C}_t^s = \mathbf{C}_t + \mathbf{B}_t (\mathbf{C}_{t+1}^s - \mathbf{R}_{t+1}) \mathbf{B}_t'$
    \EndIf
\EndFor

\end{algorithmic}
\end{algorithm}

\clearpage

\bibliography{wileyNJD-APA}%


\end{document}